\documentclass[11pt]{article}\usepackage[]{graphicx}\usepackage[]{xcolor}
\makeatletter
\def\maxwidth{ %
  \ifdim\Gin@nat@width>\linewidth
    \linewidth
  \else
    \Gin@nat@width
  \fi
}
\makeatother

\definecolor{fgcolor}{rgb}{0.345, 0.345, 0.345}
\newcommand{\hlkwb}[1]{\textcolor[rgb]{0.69,0.353,0.396}{#1}}%

\usepackage{framed}
\makeatletter
 {\par\unskip\endMakeFramed%
 \at@end@of@kframe}
\makeatother

\definecolor{shadecolor}{rgb}{.97, .97, .97}
\definecolor{messagecolor}{rgb}{0, 0, 0}
\definecolor{warningcolor}{rgb}{1, 0, 1}
\definecolor{errorcolor}{rgb}{1, 0, 0}
\newenvironment{knitrout}{}{} % an empty environment to be redefined in TeX

\usepackage{alltt} \usepackage{fullpage}
\pdfoutput=1

\newcommand\MainResultsTable{1} %% Model comparisons by log-likelihood, evaluated by a block particle filter.
\newcommand\MainFigModelDiagram{1} %% A flow diagram for the SEAIR metapopulation model.
\newcommand\MainFigSimulations{2} %% daily case report time series for models M1 and M6
\newcommand\mainFigIBPF{3} %% A flow diagram for an iterated block particle filter. 

\newcommand\suppSecOurModel{S1}
\newcommand\suppSecMobility{S1.2}
\newcommand\suppSecLiModel{S1.3}
\newcommand\suppSecBenchmark{S2}
\newcommand\suppSecReview{S3} %% Review of inference methods for metapopulation models
\newcommand\suppSecIBPF{S4}  %% The block particle filter and iterated block particle filter
\newcommand\suppSecUnconstrained{S6} %% profiles for the unconstrained li23 model
\newcommand\suppSecInference{S7} %% now profile likelihood CIs
\newcommand\suppSecResiduals{S8} %% now anomaly analysis

\newcommand\suppTableResults{S1}
\newcommand\suppFigFilterComparison{S7} %% Comparing filters on simulated data from model li23

\newcommand\myTitle{Inference on spatiotemporal dynamics for networks of biological populations}
\newcommand\myAuthorList{Jifan Li, Edward L. Ionides, Aaron A. King, Mercedes Pascual and Ning Ning}
  
\usepackage{tikz}
\usetikzlibrary{positioning}
\usetikzlibrary {arrows.meta}
\usetikzlibrary{shapes.geometric}

\newcommand\LiMobility{M$_1$}
\newcommand\LiParams{M$_2$}
\newcommand\BenchmarkIID{M$_3$}
\newcommand\BenchmarkAR{M$_4$}
\newcommand\RevisedModelUnconstrained{M$_5$}
\newcommand\RevisedModelConstrained{M$_6$}

\usepackage{enumerate,alltt,xstring}
\usepackage[ruled,noline,linesnumbered]{algorithm2e}
\SetKwFor{For}{for}{do}{end}
\SetKwFor{While}{while}{do}{end}
\SetKwInput{KwIn}{input}
\SetKwInput{KwOut}{output}
\SetKwInput{KwCplx}{complexity}
\SetKwInput{KwIndices}{note}
\SetKwBlock{Begin}{procedure}{}
\DontPrintSemicolon

\newcommand\data[1]{#1^*}
\newcommand\giventh{{\hspace{0.5mm};\hspace{0.5mm}}}
\newcommand\seq[2]{{#1}\!:\!{#2}}
\newcommand\mydot{{\,\cdot\,}}
\newcommand\given{{\,\vert\,}}
\newcommand\normal{{\mathrm{Normal}}}
\newcommand\vmeasure{V}
\newcommand\emeasure{e}

\newcommand\Time{N}
\newcommand\ttime{n}
\newcommand\Np{J}
\newcommand\np{j}
\newcommand\altNp{q}

\newcommand\prob{\mathrm{Prob}}

\usepackage[utf8]{inputenc}
\usepackage[T1]{fontenc}
\usepackage{amsmath}
\usepackage{amsfonts}
\usepackage{amssymb}
\usepackage[version=4]{mhchem}
\usepackage{stmaryrd}
\usepackage{mathtools}
\usepackage{graphicx}
\usepackage{float}
\usepackage{subfigure}
\usepackage{xurl}
\usepackage{makecell}
\usepackage{multirow}
\usepackage{multicol}
\usepackage[ruled,noline,linesnumbered]{algorithm2e}
\usepackage[normalem]{ulem}% to use \sout in feedback commands
\usepackage{bm}
\usepackage[mathscr]{euscript}
\usepackage{amsthm}
\usepackage{bbm} % for \mathbbm{1}

\newcommand\lik{L}
\newcommand\loglik{\ell}
\newcommand\myvec[1]{{\mathbf{#1}}}
\newcommand\proglang[1]{\textsf{#1}} 
\newcommand\code[1]{\texttt{#1}}
\newcommand\pkg[1]{\code{#1}}
\newcommand\class[1]{\code{#1}}
\newcommand\unit{u}
\newcommand\altUnit{\tilde{u}}
\newcommand\Unit{U}     
\newcommand\rate{\mu}
\usepackage{upgreek}
\newcommand\relativeTransmission{\upmu}
\newcommand\diagnosisDelay{T_d}
\newcommand\latency{Z}
\newcommand\infectiousPeriod{D}
\newcommand\reportRate{\alpha}
\newcommand\Rzero{\mathcal{R}_0}
\newcommand\mobilityFactor{\vartheta}
\newcommand\eqsep{\hspace{1mm}}
\newcommand\pop{P}
\newcommand\EzeroDescription{Initial $E$ in Wuhan}
\newcommand\AzeroDescription{Initial $A$ in Wuhan}
\newcommand\alphaDescription{Reported fraction}
\newcommand\before{\mathrm{be}}
\newcommand\after{\mathrm{af}}

\newcommand\E{\mathbb{E}}
\newcommand\Var{\mathrm{Var}}
\newcommand\var{\Var}

\usepackage{times}
\usepackage{url}
\usepackage{enumitem}
\usepackage[hidelinks,colorlinks=true,linkcolor=blue,citecolor=blue]{hyperref}

\usepackage{tikz,pgfplots}
\pgfplotsset{compat=1.16}
\usepgfplotslibrary{fillbetween}
\usetikzlibrary{patterns}
\usetikzlibrary{arrows,chains,matrix,positioning,scopes,fit,decorations.pathreplacing}
\tikzset{join/.code=\tikzset{after node path={%
\ifx\tikzchainprevious\pgfutil@empty\else(\tikzchainprevious)%
edge[every join]#1(\tikzchaincurrent)\fi}}}
\tikzstyle{labeled}=[execute at begin node=$\scriptstyle,
   execute at end node=$]
\newcommand{\FixedLengthArrow}{1,0}

\usepackage{color}
\definecolor{orange}{rgb}{1,0.5,0}

\definecolor{green}{rgb}{0,0.5,0}

\definecolor{purple}{rgb}{0.5,0,1}

\definecolor{teal}{rgb}{0,0.5,0.8}

\newcommand\stdError[1]{}
\newcommand\includeRates[1]{}

\newcommand\arxiv[2]{#1}  %% for arxiv version, also used for other versions

\usepackage[sort&compress,numbers]{natbib}

\title{\myTitle}

\author
{
\vspace{2mm}
\\
Jifan Li,$^{1}$ Edward L. Ionides,$^{2\ast}$ Aaron A. King$^3$, Mercedes Pascual$^4$, Ning Ning$^{1\ast}$\\
\vspace{5mm}
\\
\normalsize{$^{1}$Department of Statistics, Texas A\&M University}
\\
\normalsize{$^{2}$Department of Statistics, University of Michigan}
\\
\normalsize{$^{3}$Department of Ecology \& Evolutionary Biology, University of Michigan}
\\
\normalsize{$^{4}$Department of Environmental Studies, New York University}
\\
\normalsize{$^\ast$To whom correspondence should be addressed; E-mail: ionides@umich.edu, patning@tamu.edu}
}

\date{}

\IfFileExists{upquote.sty}{\usepackage{upquote}}{}
\begin{document}

%\baselineskip24pt

\maketitle 

\arxiv{}{

\vspace{10mm}

\noindent{\bf Classification}: physical sciences (applied mathematics) and biological sciences (ecology).

\vspace{15mm}

\noindent{\bf Key words}: Markov process; partially observed; block particle filter; maximum likelihood; metapopulation dynamics; ecology; epidemiology.

\newpage

}

\begin{abstract}
Mathematical models in ecology and epidemiology must be consistent with observed data in order to generate reliable knowledge and evidence-based policy. Metapopulation systems, which consist of a network of connected sub-populations, pose technical challenges in statistical inference due to nonlinear, stochastic interactions. Numerical difficulties encountered in conducting inference can obstruct the core scientific questions concerning the link between the mathematical models and the data. Recently, an algorithm has been developed which enables effective likelihood-based inference for the high-dimensional partially observed stochastic dynamic models arising in metapopulation systems. The COVID-19 pandemic provides a situation where mathematical models and their policy implications were widely visible, and we use the new inferential technology to revisit an influential metapopulation model used to inform basic epidemiological understanding early in the pandemic. Our methods support self-critical data analysis, enabling us to identify and address model limitations, and leading to a new model with substantially improved statistical fit and parameter identifiability. Our results suggest that the lockdown initiated on January 23, 2020 in China was more effective than previously thought. We proceed to recommend statistical analysis standards for future metapopulation system modeling.
\end{abstract}

\arxiv{}{

\begin{center}

\vspace{2mm}

\small

{\bf Significance statement}

%\section*{Significance statement}

\vspace{3mm}

\parbox{5.75in}{
{\hspace{5mm}}Complex nonlinear models are used to study spatiotemporal biological population dynamics in ecological and epidemiological contexts.
To guide evidence-based research and policy, statistical inferences must be obtained about the extent to which these models provide a quantitative description of available data.  
The necessity of doing this, and practical difficulties involved, were showcased by the COVID-19 pandemic.
We develop a statistically principled data analysis framework based on a newly developed Monte Carlo algorithm, called the iterated block particle filter.
This approach reveals strengths and weaknesses of a nonlinear stochastic partially observed mechanistic model for interacting populations that was influential early in the COVID-19 pandemic.
Our methodology, implemented in open-source software, provides a foundation for future model-based investigations of spatiotemporal population dynamics.
}
\end{center}

}

\section*{Introduction}
\label{sec:intro}

Biological populations may be structured into a collection of densely-populated communities separated by sparsely populated regions.
The network of communities, which may be cities in a human context, comprise a metapopulation.
Motivation for metapopulation modeling arises when some essential feature of the population dynamics cannot be understood from looking at a single location.
Dynamics of persistence through local extinctions and reintroductions have been extensively studied in ecology \cite{hanski98,mackenzie09}.
In epidemiology, metapopulation dynamics can be a barrier to the regional elimination and eventual eradication of a pathogen, and may determine the successful invasion of a new pathogen or a new strain of an existing pathogen \cite{metcalf21}.
In other situations, spatiotemporal dynamics may be an unavoidable component of the system under study without being the focus of the investigation \cite{zhang22,wheeler23}.

Mathematical models for biological systems are also used to inform public policy, despite delicate issues in their implementation and interpretation \cite{saltelli20}.
Indeed, it can be practically impossible to make sense of the nonlinear stochastic interactions driving biological dynamics without representing them via a model \cite{mccabe21,lofgren14}.
However. both operational and conceptual difficulties arise when developing these models.
Operationally, we seek to fit complex models using statistically valid, reproducible and transparent methods.
Conceptual difficulties arise when drawing causal conclusions from fitting models to observational data, giving rise to opportunities for incorrect conclusions due to missing variables or other forms of model misspecification.
A model assimilated to data guarantees that assumptions have been framed in a way consistent with certain facts, and evidence for predictive skill can support the value of the model construction.

A recent growth in the study of metapopulation dynamics has been driven partly by the COVID-19 pandemic \cite{li20,wu20,wang22,prieto22,cascante-vega22,pizzuti20,alleman21,yang21,engebretsen23} and in part by methodological advances facilitating the fitting of metapopulation models to spatiotemporal data.
Until the start of this millennium, developing dynamic models with both statistical and scientific justification was a longstanding open problem for even a single community \cite{bjornstad01}.
Over the past two decades, new algorithms \cite{ionides06,toni09,andrieu10,ionides15} and software \cite{king16,kristensen16,devalpine17}, together with ever-increasing computational resources, have enabled routine inference for low-dimensional nonlinear partially observed stochastic dynamic systems.
However, fundamental algorithmic scalability issues known as the ``curse of dimensionality'' lead to difficulties with the high-dimensional systems arising in metapopulation inference.
These issues are clearest for Monte Carlo techniques based on importance sampling \cite{bengtsson08} but are also evident in the need for variational approximations for large Monte Carlo Markov Chain (MCMC) calculations \cite{blei17}. 
Thus, data analysis for metapopulation models has lagged behind the analysis of low-dimensional time series data for biological dynamics.
Recent developments enable this gap to be closed, as we demonstrate via a reanalysis of COVID-19, viewed from the context of the ability to draw evidence-based scientific conclusions about the dynamics of the emerging pandemic in January and February 2020.

Biological systems are characterized by nonlinear stochastic dynamics together with incomplete and noisy measurements  \cite{bjornstad01}.
We therefore focus on the class of partially observed Markov process (POMP) models \cite{breto09}, acknowledging that deterministic models can be conceptually useful but are problematic as statistical explanations of noisy systems \cite{king15,wheeler23}.
The Markov property asserts that the dynamic process has no memory conditional on its current state, which is algorithmically convenient while being scientifically nonrestrictive since we can choose what to include in the state.
Metapopulation models consider a multivariate system state at each location and so we require methods tailored for high-dimensional POMP models.
Simplifications arise if models and data are limited to binary presence-absence, or a small discrete set of values at each location \cite{mackenzie09}, but we are concerned with situations where time series of abundance data are available, such as case reports for infectious diseases.
We focus on two inferential approaches for high-dimensional POMP models, the block particle filter (BPF) and the ensemble Kalman filter (EnKF).
Other alternatives are reviewed in Supplementary Sec.~{\suppSecReview}.

EnKF was developed in the context of massive geophysical models.
It combines an ensemble representation of the latent state with a computationally efficient update rule inspired by the scalable linear Kalman filter, providing an approach with excellent scalability \cite{evensen09book,evensen22}.
For biological systems, EnKF was first demonstrated as a computationally convenient tool for compartment models at a single location \cite{shaman12,yang14}.
Subsequently, it has been applied for epidemiological metapopulation inference \cite{li20,kramer20-plos-cb}.
However, the linearization in the EnKF filter update rule can be problematic for highly nonlinear systems \cite{evensen22,ionides23-jasa}.
Further, a linear update rule is not appropriate for small, discrete populations unless EnKF is embedded within a MCMC algorithm \cite{katzfuss19}.
By contrast, particle filter methods \cite{doucet11} avoid linearization and are directly applicable to discrete and continuous latent states.

For low-dimensional systems, particle filter methods are broadly applicable; they permit consideration of arbitrary nonlinear dynamics and require the model to be specified only via a simulator \cite{breto09,king16}.
Particle filters enable statistically efficient use of data, since they provide an evaluation of the likelihood function required for Bayesian or likelihood-based inference, with approximation resulting only from finite Monte Carlo effort.
For high-dimensional systems, scalability considerations demand further approximations since particle filters suffer acutely from the ``curse of dimensionality'' \cite{bengtsson08}.
The BPF algorithm modifies the particle filter to achieve scalability by carrying out local resampling on spatial neighborhoods known as blocks.
This avoids the linear update rule used by EnKF \cite{rebeschini15}.
It is an empirical question whether the different approximations inherent in EnKF and BPF are successful on metapopulation models, with prior evidence favoring BPF \cite{ionides23-jasa}.
In the following example, we demonstrate that BPF can be effective for a practical metapopulation data analysis.
We show that the resulting likelihood-based inference framework provides opportunities for model criticism, leading to rigorous assessment of model fit and improved advice on public policy decisions.

\section*{Metapopulation analysis of COVID-19 spread in China}
\label{sec:covid}

We reconsider the influential analysis of COVID-19 from early in the pandemic by Li et al. \cite{li20}.
This analysis provided estimates of transmission parameters and the effect of the lockdown in China using the limited data available at the time.
Other teams have fitted models to address similar questions \cite{kraemer20,yang21,brett23} but the study by \cite{li20} is distinctive for fitting a stochastic mechanistic metapopulation model to extensive spatiotemporal data.
The results were published in May, 2020, based on reported cases from January 10 to February 8 of that year.
The state-of-the-art spatiotemporal analysis was possible on an urgent timescale because the team of researchers had developed their methodology in a sequence of previous situations \cite{shaman12,yang14,yang15,pei18}.
The paper is written with attention to reproducibility, and the main results are strengthened by various supporting analyses in an extensive supplement.
While examining the points mentioned above, we have identified various limitations that could have been mitigated by adhering to the aforementioned recommendations.
Our goal is not to criticize any specific paper, but rather to build on the timely analysis of \cite{li20} to demonstrate how recently developed techniques provide possibilities to carry out improved data analysis in future.

\usetikzlibrary{positioning}
\usetikzlibrary {arrows.meta}
\usetikzlibrary{shapes.geometric}

\begin{figure}
\begin{center}
%%%%% SEAIR diagram
\resizebox{!}{6cm}{
\begin{tikzpicture}[
  square/.style={rectangle, draw=black, minimum width=0.8cm, minimum height=0.8cm, rounded corners=.1cm, fill=blue!8},
  travel/.style={circle, draw=black, minimum width=0.85cm, minimum height=0.8cm, fill=green!8},
  report/.style={shape=regular polygon, regular polygon sides=8, draw, fill=red!8,minimum size=0.9cm,inner sep=0cm},
  bendy/.style={bend left=25},
  >/.style={shorten >=0.25mm}, % redefine arrow to stop short of node
  >/.tip={Stealth[length=1.5mm,width=1.5mm]} % redefine arrow style
]
\tikzset{>={}}; % this is needed to implement the arrow redefinition
\tikzset{every path/.style={line width=0.8 pt}}

\draw[rounded corners, green, thick] (-1.2, 1) rectangle (5.2, 2.5) {};
\draw[rounded corners, blue, thick] (-1.2, -2.4) rectangle (7.7, 0.7) {};
\draw[rounded corners, red, thick] (3.75, -4.15) rectangle (10.2, -2.85) {};
\node (between) at (5.4,2.3) [green, anchor=north west, font=\large] {Travel between cities};
\node (within) at (7.9,0.5) [blue, anchor=north west, font=\large] {Disease dynamics within cities};
\node (report) at (10.4,-3.05) [red, anchor=north west, font=\large] {Reported cases};

\node (S) at (-0.5,0) [square] {S$_u$};
\node (E) at (2,0) [square] {E$_u$};
\node (A) at (4.5,0) [square] {A$_u$};
\node (I) at (4.5,-1.75) [square] {I$_u$};
\node (R) at (7,0) [square] {R$_u$};

\node (T1) at (-0.5,1.75) [travel] {T};
\node (T2) at (2,1.75) [travel] {T};
\node (T3) at (4.5,1.75) [travel] {T};

\node (Ca) at (4.5,-3.5) [report] {C$^a_u$};
\node (Cb) at (7,-3.5) [report] {C$^b_u$};
\node (C) at (9.5,-3.5) [report] {C$_u$};

\node (V1) at (2.5,-1.75) {};
\node (V2) at (3,-3.5) {};
\draw [->] (E) -- (Ca -| V2) -- (Ca);
\draw [->] (I -| V1) -- (I);

\draw [->] (S) -- (E);
\draw [->] (A) -- (R);

\draw [->] (Ca) -- (Cb);
\draw [->] (Cb) -- (C);

\draw [->] (I.east) -- (R);
\draw [->] (E) -- (A);

\draw [->, bendy] (S) to (T1);
\draw [->, bendy] (T1) to (S);

\draw [->, bendy] (E) to (T2);
\draw [->, bendy] (T2) to (E);

\draw [->, bendy] (A) to (T3);
\draw [->, bendy] (T3) to (A);

\end{tikzpicture}
}
\end{center}
\vspace{-5mm}
\caption{A flow diagram for the SEAIR metapopulation model.
Each individual in city $u$ is a member of exactly one of the square blue compartments.
Individuals entering the reportable infectious compartment, $\mathrm{I}_u$ for city $u$, are simultaneously included in the delayed reporting process compartment, $\mathrm{C^a_u}$.
Upon arrival at the final reporting compartment, $\mathrm{C}_u$, the individual is included in the case report for city $u$.
Individuals in $\mathrm{A}_u$ are not reportable and transmit at a reduced rate.
Movement of individuals between cities occurs by transport to and from a transport compartment, $\mathrm{T}$.
The number of individuals moving between each pair of cities is based on 2018 data from Tencent.
Movement is modeled only for susceptible, exposed, and undetected infections.}\label{fig:flow_diagram}
\end{figure}
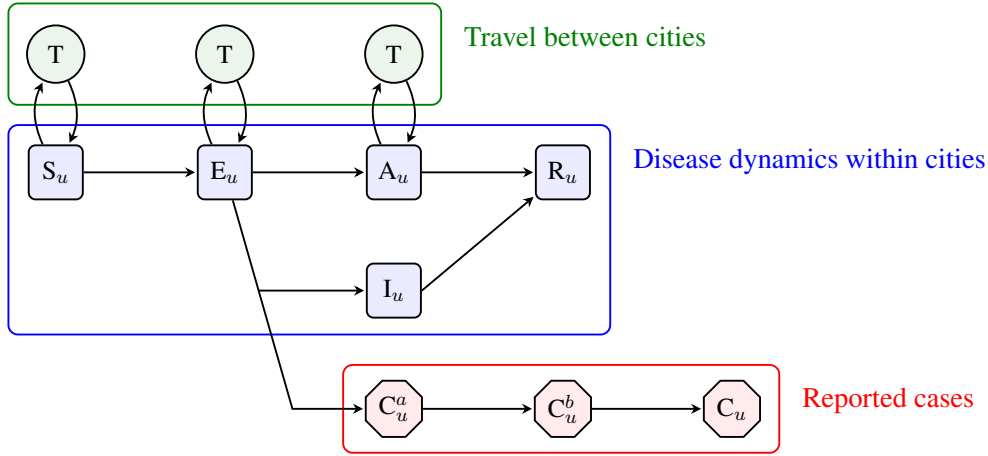

For our metapopulation system, the sub-populations are 373 provincial cities in China (meaning cities with administrative responsibility for an entire region) and the data are daily reported COVID-19 cases. 
COVID-19 dynamics are represented by a Susceptible Exposed Asymptomatic Infectious Recovered (SEAIR) epidemic model. 
Questions of urgent interest early in the pandemic include the relative transmissibility of reported to unreported cases, the fraction of unreported cases, and the effect on transmission of movement restrictions imposed on and around January 23 \cite{li20}.
The model structure is illustrated by the flow diagram in Fig.~\ref{fig:flow_diagram}.
The Methods section provides additional description, with Eq.~(\ref{eq:muSE}) giving the modeled rate at which susceptible individuals become infected.
The complete model specification and estimated parameter values are in Supplementary Sec.~{\suppSecOurModel}.

We consider different model implementations within this structure.
Our starting point is model {\LiMobility} which is based on the model of \cite{li20} and is described in Supplementary Sec.~{\suppSecLiModel}.
We consider the full dataset, from January 10 to February 8, with transmission parameters re-estimated following the lockdown on January 23; these correspond to the periods~1 (January 10 to January 22) and 3 (January 24 to February 8) of \cite{li20}.
Some minor differences between {\LiMobility} and \citep{li20} were introduced to enable us to place their model within the general framework of spatiotemporal partially observed continuous-time Markov process models described by \citep{ionides23-jasa}.
Despite these modifications, simulations from {\LiMobility}, using the parameters of \citep{li20}, closely match simulations from the code provided by \cite{li20} (Supplementary Sec.~\suppSecLiModel).
However, inspection of the mobility data reveals that some small cities have no recorded incoming travelers, and therefore no possibility of a SARS-CoV2 introduction within {\LiMobility} (or the model of \cite{li20}) (Supplementary Sec.~\suppSecMobility).
This minor limitation formally results in a likelihood of zero for {\LiMobility} (i.e., it is impossible for the simulation model to reproduce the observed spatiotemporal dataset), and hence a log-likelihood of $-\infty$.

\begin{table}[ht]
\small
\begin{tabular}{c|rr|l}
Model & loglik & df & description
\\
\hline
{\LiMobility} &
  $-\infty$ & 11 &
  SEAIR model using the parameter values and mobility data of \cite{li20}
\\
{\LiParams} &
  -14985.0\stdError{569.8}& 11 &
  Adjusted mobility and measurement in {\LiMobility}
\\
{\BenchmarkIID} &
  -11257.9  & 374 &
  Independent identically distributed negative binomial
\\
{\BenchmarkAR} &
 -10825.3  & 375 &
  Autoregressive negative binomial
\\
{\RevisedModelUnconstrained} &
  -9088.2\stdError{3.4}&
  15 &
  Adding overdispersed dynamics to {\LiParams} and refitting
\\
{\RevisedModelConstrained} &
  -9116.5\stdError{10.1}&
  13 &
  Latent and infectious durations unchanged by lockdown in {\RevisedModelUnconstrained}.
\\

\hline
\end{tabular}
\caption{Model comparisons by log-likelihood, evaluated by a block particle filter. The degrees of freedom (df) is the number of estimated parameters. 
}
\label{tab:loglik}
\end{table}

We addressed the problematic mobility data in {\LiMobility} by adding some additional transportation based on a gravity movement model, as described in Supplementary Sec.~{\suppSecMobility}, giving rise to model~{\LiParams}.
We implemented an additional adjustment between models {\LiMobility} and {\LiParams} to align the measurement model with the ensemble Kalman filter (EnKF) inference method presented by \cite{li20}.
That EnKF implementation involved specifying a quantity called the observation error variance, defined as a function of the observed cases, to quantify the uncertainty in the measurements.
Within the POMP specification, the measurement variance can depend on the latent state but not directly on the observed data.
To interpret the choice of EnKF observation variance within the POMP framework, we specified the measurement model for~{\LiParams} to have equivalent scaling to the choice of \citep{li20}, but with dependence on the reported cases replaced by dependence on the modeled, but unobserved, exact case count.

Based on a comparison of various nonlinear spatiotemporal filters (Supplementary Fig.~\suppFigFilterComparison) we evaluated the log-likelihood for {\LiParams} using a block particle filter (Table~\ref{tab:loglik}).
To account for model overfitting, the number of estimated parameters can be subtracted from the log-likelihood to obtain a comparison equivalent to Akaike's Information Criterion (AIC) \citep{burnham02}.
When the difference in log-likelihood is large compared to the difference in degrees of freedom, the ordering of statistical goodness-of-fit is clear without presenting formal statistical hypothesis tests.

To find out whether this log-likelihood value suggests that the model is satisfactory, we compare it with two simple statistical models:
{\BenchmarkIID} simply models the daily case report for each city as an independent identically distributed (IID) negative binomial random variable; 
{\BenchmarkAR} adds an autoregressive component to {\BenchmarkIID} (see Supplementary Sec.~\suppSecBenchmark).
We see from Table~\ref{tab:loglik} that both {\BenchmarkIID} and {\BenchmarkAR} outperform {\LiParams} by many units of log-likelihood.
Likelihood can properly be compared between different models for the same data, with statistical uncertainty in log-likelihood differences arising on the unit scale \citep{pawitan01}.
When the fit of a mechanistic model is inferior to a simple statistical model, we learn that the mechanistic model has room for improvement as a description of the data, but we do not immediately learn what the deficiency is.
The development of methods for formal statistical fitting of mechanistic models has led to increased understanding of the importance of appropriate modeling of over-dispersed variation in the stochastic dynamics \cite{he10,stocks20,whitehouse23}.
We therefore hypothesized that the fit of {\LiParams} could be improved by permitting additional dynamic noise.

A standard way to convert a deterministic model, constructed as a system of ordinary differential equations, into a stochastic model is to reinterpret the rates of the deterministic system as rates of a Markov chain \citep{keeling09}.
This places limits on the mean-variance relationship of the resulting stochastic model \citep{breto11}.
Models allowing greater variability that permitted by this construction are said to be over-dispersed.
We added multiplicative white noise to the transmission rate, following the approach of \citep{breto09,he10}, giving rise to model~{\RevisedModelUnconstrained}.
We fitted the model using an iterated block particle filter to maximize the likelihood \citep{ionides22,ning23}.
The block filter approximation has also proven helpful for spatiotemporal inference when using alternatives to particle filtering and alternatives to maximization by iterated filtering \cite{whitehouse23}.
In the current context, the block particle filter was found to be more effective for likelihood evaluation than a test suite of alternative filters including the ensemble Kalman filter (Supplementary Fig.~\suppFigFilterComparison).
The iterated block particle filter maximizes the block particle filter likelihood using an iterated filtering algorithm \citep{ionides15} adapted to the structure of a block particle filter.

Table~1 shows that model~{\RevisedModelUnconstrained} outperforms simple statistical benchmarks, obtaining a competitive likelihood with relatively few parameters.
From a statistical perspective, {\RevisedModelUnconstrained} is therefore an adequate statistical description of the data.
However, some parameters of {\RevisedModelUnconstrained} were weakly identified by the data, especially in the pre-lockdown time interval within which there were relatively few reported cases (Supplementary Sec.~\suppSecUnconstrained).
When the evidence about the model parameters in the data is weak, the ambiguity may be resolved by other, unmodeled and poorly understood, aspects of the data.
This risks leading to undesirable situations where substantial conclusions about questions of interest could be driven by the weaknesses of the model rather than its strengths.
In Supplementary Sec.~{\suppSecUnconstrained}, we show how the flexibility of {\RevisedModelUnconstrained} can be used to obtain a high likelihood via an unplausibly long estimated duration of infection during the pre-lockdown period, with the estimate suddenly reducing after lockdown.
 We resolved this issue by constraining the latent and infectious periods to be the same before and after lockdown, leading to model {\RevisedModelConstrained}.
 The additional constraints of {\RevisedModelConstrained} lead to a small loss of likelihood compared to {\RevisedModelUnconstrained}, but the fit remains competitive compared to the benchmark models, and the stronger identifiability facilitates the interpretation of estimated parameters.
Calculating the log-likelihood for each model in Table~\ref{tab:loglik} requires extensive computation to produce a single number which contains essentially all the information about the statistical fit of the model.
However, deeper investigation is required to understand what characteristics of the models and data causes the differences in these numbers, and the practical consequences of the numerical results.
As a starting point, Fig.~\ref{fig:panel_plot} plots the data next to simulations from models {\LiMobility} and {\RevisedModelConstrained}.
Visually, the comparison confirms {\RevisedModelConstrained} as a reasonable representation of the data.
Both {\LiMobility} and {\RevisedModelConstrained} overestimate cases before day 14, but the context of rapidly increasing awareness and growing diagnostic capabilities is hard to quantify.

\begin{knitrout}
\definecolor{shadecolor}{rgb}{0.969, 0.969, 0.969}\color{fgcolor}\begin{figure}

{\centering \includegraphics[width=6.5in]{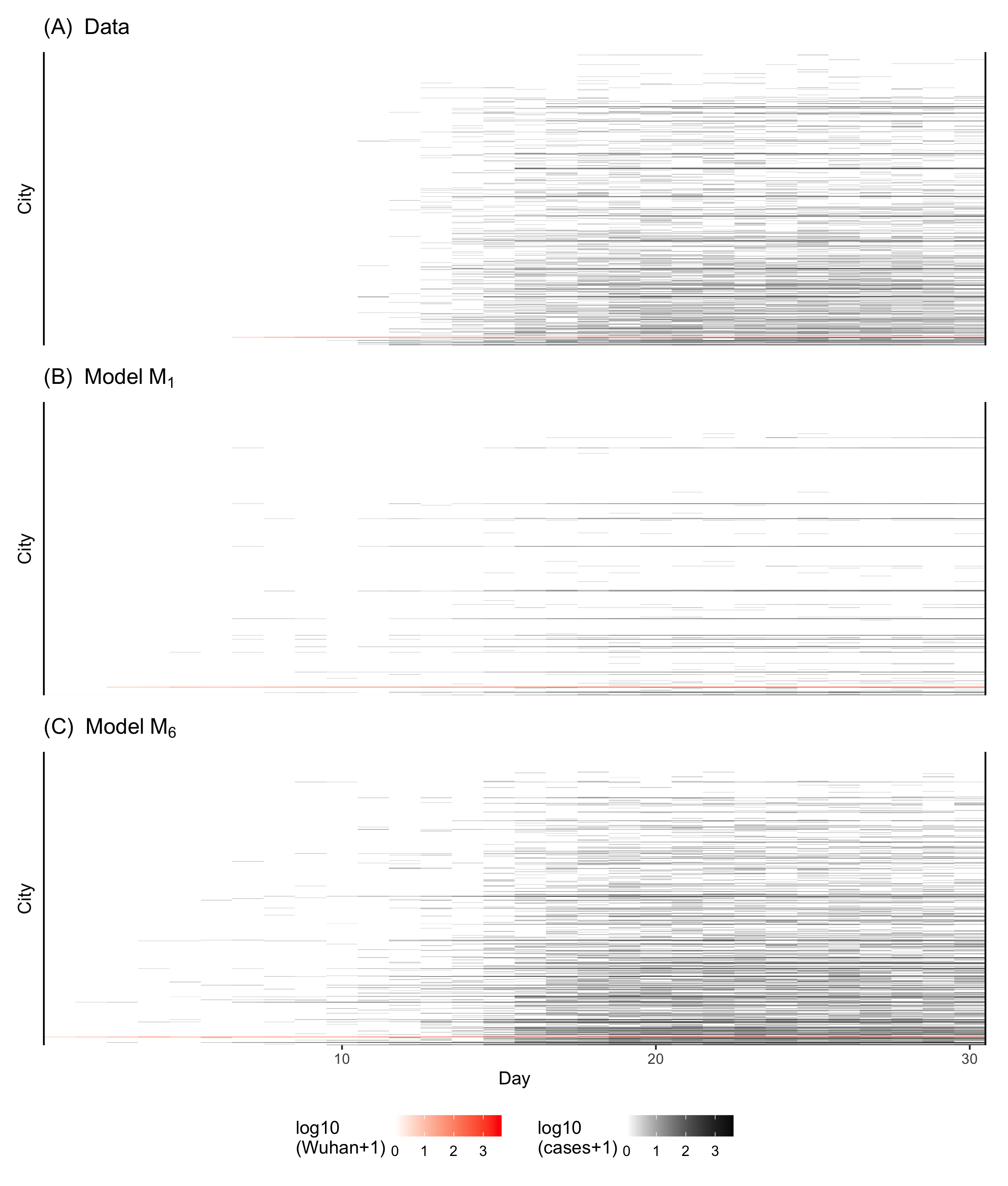} 

}

\caption{Daily case report time series for 373 cities: (A) the real data; (B) a simulation from model {\LiMobility}; (C) a simulation from model~{\RevisedModelConstrained}. Within each panel, cities are ordered by population, largest on the bottom row.}\label{fig:panel_plot}
\end{figure}

\end{knitrout}

Parameter values for models~ {\LiMobility},~{\RevisedModelUnconstrained} and~{\RevisedModelConstrained} are reported in Supplementary Table~{\suppTableResults}.
Here, we discuss the estimated basic reproductive number (i.e., the expected number of secondary infections from one index case in a fully susceptible population), denoted by $\Rzero^{\before}$ and $\Rzero^{\after}$ before and after the January 23rd lockdown.
$\Rzero$ is calculated by the formula in Eq.~(\ref{eq:Rzero}).
Our estimates for model {\RevisedModelConstrained}, are $\Rzero^{\before}=3.51$ with confidence interval (CI) (3.31,3.72) and $\Rzero^{\after}=0.70$  with CI (0.65,0.77), where the estimates and their associated 95\% CIs obtained by profile likelihood (Supplementary Sec.~\suppSecInference).
This implies that the Chinese government non-pharmaceutical interventions instituted on and around January 23 reduced $\Rzero$ by a factor of $5.0$.
By contrast, the estimates of \cite{li20} are $\widetilde\Rzero^{\before}=2.38$  with CI $(2.03,2.77)$ and $\widetilde\Rzero^{\after}=0.98$  with CI $(0.83,1.16)$, implying reduction by a factor of $2.4$.
For comparison, interventions implemented across a panel of 41 countries (34 European) were estimated to reduce $\Rzero$ by a factor of $4.3$  with CI $(2.9,6.7)$ \citep{brauner21}.
Our estimate for $\Rzero$ before lockdown is toward the high end of previous estimates based on data up to February 2020, reviewed by \citep{park20-jcm}.
An alternative metaopopulation analysis of the pre-lockdown China data, with a deterministic transmission model, obtained an $\Rzero$ estimate of 3.11  with CI  (2.39,4.13) \citep{read21}.
Our $\Rzero$ estimate is consistent with pre-lockdown estimates from other locations, such as New York city, for models that include asymptomatics \cite{subramanian21}.

The likelihood-based confidence intervals for {\RevisedModelConstrained} are narrower than the intervals from \cite{li20}.
However, {\RevisedModelConstrained} fits two fewer parameters than {\RevisedModelUnconstrained}, and the latter is more directly comparable to the model of \cite{li20}.
For {\RevisedModelUnconstrained}, the likelihood-based analysis leads to some wide confidence intervals, revealing the weakly identified parameters.

Our model inherits the property of \citep{li20} that infections arising during the pre-lockdown period will generally be reported during the lockdown, due to the reporting delay modeled as a distributed delay with a mean of 9 days pre-lockdown and 6 days post-lockdown.
Thus, the model is permitted to explain the data by inferring rapid, unreported spread prior to January 23.
Despite this shared constraint on the form of the model, conclusions of our analysis differ from \cite{li20}.
Beyond the estimates of $\Rzero$, a notable difference is that we find the estimated transmissibility of observed cases is close to that of unobserved cases, especially before lockdown (Supplementary Table~S1).

Not all models are equal, and we have demonstrated an approach which evaluates the extent to which the postulated models statistically explain the observed data. 
Our analysis cannot disprove the possibility of an alternative model which explains the data even better via an alternative model specification, perhaps leading to alternative conclusions.
Indeed, our methods are designed to facilitate others to develop and demonstrate superiority to our own analysis when such advances are available.

If a mechanistic model has likelihood competitive with statistical benchmarks, it is anticipated to have simulations that are qualitatively comparable to the data.
Since the model specification is inevitably imperfect, and is accounted for in the model fitting by noise processes, we expect simulations from the fitted model to have somewhat more stochastic variation than the data.
By contrast, models which are structurally unable to provide sufficient variability to explain the data must give rise to simulations with too little stochasticity (as shown in Fig.~\ref{fig:panel_plot}).
Models that have simulations with implausibly little variability give rise to claims of excessive confidence about the uncertainty surrounding estimated parameters.
This phenomenon may be clearest when CIs are calculated using  parametric bootstrap approach, involving re-estimation of parameters using artificial datasets simulated from a fitted model.
However, it also applies for classical CI and Bayesian credible interval constructions.
Thus, CIs from mechanistic models that outperform statistical benchmarks are anticipated to be conservative, whereas CIs from models with insufficient variability to explain the data are generally anti-conservative.
Requiring model likelihoods to be comparable to statistical benchmarks therefore improves the credibility of uncertainty intervals as well as improving the accuracy of point estimates.

\section*{Discussion}
\label{sec:conclusion}

Advances in statistical methodology will drive an increase in the number of spatiotemporal models fitted to epidemiological data.
Our research demonstrates that techniques proven effective in low-dimensional systems, such as population dynamics at one or two locations, can be extended to address larger metapopulation systems.
This extension allows us to leverage well-established best practices from time series analysis, leading to a statistically principled approach. This approach enables us to identify and rectify model limitations that might otherwise remain undetected.
Failure to address these weaknesses can lead to issues of irreproducibility and the provision of suboptimal policy recommendations when developing models for complex dynamic systems \cite{saltelli20,ioannidis22}.
Principles of good data analysis for population dynamics are presumably similar to general principles of data science \cite{yu20} but require some adaptation to the specific situation.
Here, we build on \cite{saltelli20,ioannidis22,yu20}, by demonstrating the feasibility and desirability of metapopulation analysis meeting the specific set of criteria outlined below.

\begin{enumerate}

\item \label{point:i} {\bf Likelihood-based statistical inference}. A model, in conjunction with data, defines a likelihood function that quantifies the goodness of fit of the model and the data for each parameter value.
For mechanistic models, it is usually impossible to write down the likelihood explicitly, but it still exists implicitly.
Such methods extract all available information in the data about model parameters \cite{pawitan01}.
Log-likelihood is also a proper scoring rule for comparing probabilistic forecasts \cite{gneiting07} and therefore provides a sensitive objective tool for model selection and identification of model misspecification.
Whereas cross-validation and out-of-sample fit are standard benchmarks in machine learning settings \cite{yu20}, likelihood is better suited to situations with relatively small, spatiotemporally structured datasets.
Likelihood-based inference via particle filters has been considered inaccessible for metapopulation models due to the ``curse of dimensionality'' \cite{bengtsson08}.
However, block particle filter methods can be effective on metapopulation models, as demonstrated in this paper and previously \cite{ionides23-jasa,ning23,wheeler23}.
All high-dimensional nonlinear filters entail numerical approximation, and these can be assessed by comparing predictive skill (i.e., the estimated log-likelihood) between different filters.
The ensemble Kalman filter provides a suitable point of comparison, since it has excellent scalability properties, modest capability to handle nonlinearities, and has been demonstrated on various epidemiological systems \cite{shaman12,yang14,yang15,pei18,yang21,kramer20-plos-cb,cascante-vega22}.

\item \label{point:ii} {\bf Statistical benchmarks}.
The challenge of fitting intricate nonlinear models to extensive datasets makes it difficult for researchers to evaluate the limitations of their models and methods.
Readers also can struggle to determine whether the proposed model has been adequately tested.
It is therefore advisable to incorporate benchmarks for evaluating model performance in comparison to relatively simple statistical models \citep{he10}.
This approach helps determine whether complex models provide a satisfactory level of explanatory power.
In the first instance, these benchmarks can be applied to the entire dataset; subsequent analysis can focus on dissecting the contributions from various subsets of the data to gain a comprehensive understanding of which parts of the data drive the overall assessment.
The likelihood provides a suitable quantity for comparison between different models for the same data \cite{pawitan01}.
If we find a simple statistical model with a log-likelihood many units higher than a mechanistic model, we have discovered that the mechanistic model is unable to explain some substantial aspect of the data.
At the very least, this discrepancy should be identified and discussed.

\item \label{point:iii} {\bf Residual analysis}.
Introductory statistics classes, when covering linear regression, emphasize that a careful and complete data analysis involves examining deviations from the fitted model \cite{faraway14}.
This is typically achieved by plotting residuals, a suitably rescaled measure of disparities between each observation and its corresponding fitted value.
A relevant measure of residual in the current context is the {\it log-likelihood anomaly}, defined as the discrepancy between the mechanistic fit and a benchmark for components of the likelihood at each observation.
Supplementary Sec.~{\suppSecResiduals} describes how these tools were used for developing and evaluating model~{\RevisedModelConstrained}.

\item \label{point:vii} {\bf Uncertainty}.
Reliable conclusions should be robust to plausible variations in data, models, and algorithms \cite{yu20}.
Standard statistical methods provide measures of uncertainty, and the validity of these measures depends critically on the statistical validity of the model.
Appropriate modeling of overdispersion can be critical to accurate assessment of uncertainty for dynamic models \citep{breto09,he10,stocks20,whitehouse23}.

\item \label{point:viii} {\bf Reproducibility and extendability}.
Observational studies are not generally replicable in an experimental sense.
However,  the numerical conclusions should be readily reproducible from the observations.
A substantial part of the value of a computational model is that it permits {\it in silico} experimentation of the modeled system.
The authors should build and share a computational environment that not only reproduces published numbers but also facilitates future {\it in silico} experimentation.
Subsequent research should readily be able to challenge the assumptions of the model in light of subsequent data.
In practice, this requires that the scientists provide a free, open-source software environment within which the published analysis can readily be reproduced, modified and extended \citep{gentleman07,wheeler23}. 
Development of a principled data analysis environment assists the researchers to explore their own models and data, and this environment should be shared as part of the publication process.
In practice, this involves encapsulating data analysis within a software package that immerses the user in a documented environment where the models, methods and data used for the article can be readily be experimented with.
Trustworthy data analysis should be supported by unit testing and documentation, and the quality of this support should be one of the considerations in evaluation of the data analysis.
In other words, the article presenting the research should be part of a compendium \cite{gentleman07}.
The compendium for this article is comprised of the article source code, at \url{https://github.com/jifanli/metapop_article}, together with the software environment for the data analysis, provided by the R package at \url{https://github.com/jifanli/metapoppkg}.

\item \label{point:ix} {\bf Appropriate conclusions from observational data}.
In the absence of a randomized controlled experiment, the care required to move from a fitted model parameter to a causal claim is well known in linear regression analysis \cite{faraway14}.
The same principles apply to nonlinear dynamic metapopulation models: the model structure may be informed by prior scientific knowledge, and the model may statistically explain population-level data, yet observational data cannot readily rule out the possibility of alternative explanations.
A model may be called hypothetically causal when it is consistent with scientifically plausible causal mechanisms, but the model fitting process does not itself validate these assumptions---this is a common situation for metapopulation modeling.

\end{enumerate}

In conclusion, the study of metapopulation dynamics will continue to benefit from advances in algorithms, software, and data analysis methodologies.
The models should undergo critical scrutiny to delineate their strengths and weaknesses, following evaluation procedures such as we have described in this paper. 
With due care, these models can unearth limitations in existing knowledge, investigate hypotheses that may extend our knowledge, and furnish us with valuable predictive tools.

\section*{Methods}

\noindent {\it Data}.
COVID-19 case reports, city population counts, and the time-varying matrix of movement between cities, were taken from \citep{li20}.
Some erroneous numbers, revealed by our log-likelihood anomaly analysis, were subsequently modified as described in Supplementary Sec.~{\suppSecOurModel}.

\vspace{3mm}

\noindent {\it Model}.
All the mechanistic models under consideration have an SEAIR structure, as described in Figure~\ref{fig:flow_diagram}.
Supplementary Sec.~{\suppSecOurModel} provides a full mathematical representation of the SEAIR model.
Briefly, the force of infection on susceptible individuals for city $u$ due to symptomatic and asymptomatic individuals in city $u$ is given by
\begin{equation}
\label{eq:muSE}
\mu_{S_{u}E_{u}} = \beta \left(\frac{I_{u}(t)+ \relativeTransmission A_{u}(t)}{\pop_{u}(t)}\right) d\Gamma_{u}/dt,
\end{equation}
where $\beta$ is a transmission rate, $\relativeTransmission$ is the relative transmissibility of asymptomatic cases, and $\pop_u$ is the city population.
The Gamma white noise process, $d\Gamma_{u}/dt$, allows for stochastic variation in the transmission rate \cite{breto09}.
The rate at which individuals move between each pair of cities is defined by a time-varying matrix based on high-resolution Tencent data from 2018, as described in Supplementary Sec.~{\suppSecMobility}.
The basic reproductive number is given by
\begin{equation}
\label{eq:Rzero}
\Rzero=\big(\reportRate+(1-\reportRate)\relativeTransmission\big)\infectiousPeriod\beta,
\end{equation}
where $\infectiousPeriod$ is the mean infectious period and $\reportRate$ is the fraction of cases severe enough to be reported.

\begin{figure}[!ht]
\begin{center}
\begin{tikzpicture}[
  every node/.style={scale=0.85},
  square/.style={rectangle, draw=black, minimum width=5.5cm, minimum height=0.7cm, rounded corners=.1cm,font=\ttfamily},
  block/.style={rectangle, draw=black, minimum width=1.5cm, align=center, rounded corners=.1cm,font=\ttfamily},
  >/.style={shorten >=0.4mm}, % redefine arrow to stop short of node
  >/.tip={Stealth[length=2.5mm,width=1.5mm]} % redefine arrow style
]
\tikzset{>={}}; % this is needed to implement the arrow redefinition
  \node (initialize)   at (0,7.3)  [square] {Initialize model \& parameters};
  \node (perturb)    at (0,6) [square,fill=blue!8] {Perturb parameters};
  
  \node (predict)  at (0,4.7)  [square] {Predict:~stochastic dynamics};
  
  \node (reweight1) at (-3,3)  [block] {Reweight \\ Block 1};
  \node (reweight2) at (-1,3)  [block] {Reweight\\ Block 2};
  \node (reweight3) at (1,3) {\texttt{...}};
  \node (reweightK) at (3,3)  [block] {Reweight\\ Block K};

  \node (resample1) at (-3,1.5)  [block] {Resample\\state};
  \node (resample2) at (-1,1.5)  [block] {Resample\\ state};
  \node (resampleK) at (3,1.5)  [block] {Resample\\ state};

  \node (resample3) at (1,1) {\texttt{...}};

  \node (params1) at (-3,0.5)  [block, fill=blue!8] {Resample\\ params};
  \node (params2) at (-1,0.5)  [block, fill=blue!8] {Resample\\ params};
  \node (paramsK) at (3,0.5)  [block, fill=blue!8] {Resample\\ params};

  \node (recombine) at (0,-1.2)  [square] {Recombine};

  \node (N) at (5.5,3)  [draw,diamond,aspect=1.5] {\texttt{n=1:N}};
  \node (M) at (7.5,3)  [draw,diamond,aspect=1.5,fill=blue!8] {\texttt{m=1:M}};
  \node (mysouth) at (0,-1.8) {};

  \draw[->] (initialize.south) -- (perturb.north);
  \draw[->] (perturb.south) -- (predict.north);
  \draw[->] (predict) -- (reweight1.north);
  \draw[->] (predict) -- (reweight2.north);
  \draw[->] (predict) -- (reweightK.north);
  \draw[->] (reweight1) -- (resample1);
  \draw[->] (reweight2) -- (resample2);
  \draw[->] (reweightK) -- (resampleK);
  
  \draw[->] (params1.south) -- (recombine);
  \draw[->] (params2.south) -- (recombine);
  \draw[->] (paramsK.south) -- (recombine);
  \draw[->] (recombine.east) -- (recombine -| N) -- (N);
  \draw[->] (N) -- (perturb -| N) -- (perturb); 
  \draw[->] (recombine.south) -- (mysouth -| recombine.south) -- (mysouth -| M) -- (M);
 \draw[->] (M) -- (initialize -| M) -- (initialize);
\end{tikzpicture}

\end{center}
\caption{A flow diagram for an iterated block particle filter. The inner loop is a block particle filter and the outer loop enables parameter estimation.} \label{fig:ibpf}
\end{figure}
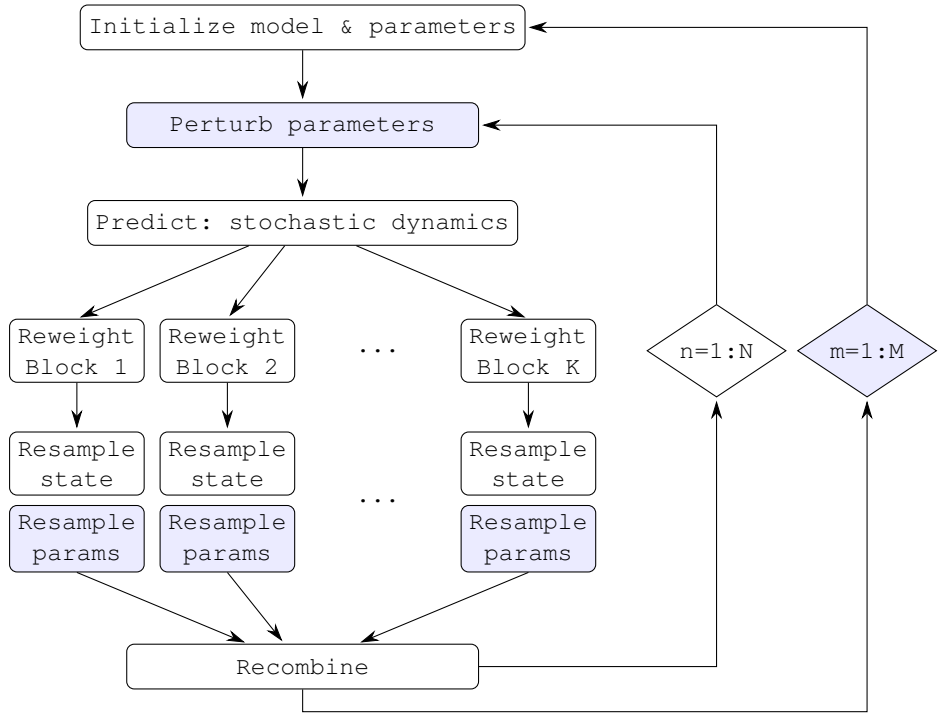

\vspace{3mm}

\noindent {\it Likelihood evaluation and maximization}.
The log-likelihood for the SpatPOMP models was calculated using BPF.
This log-likelihood was then maximized using an iterated block particle filter (IBPF) \citep{ionides22,ning23}.
A diagram representing the IBPF algorithm is shown in Figure~\ref{fig:ibpf}.
The inner loop, $n=1,\dots,N$, corresponds to BPF applied to an extension of the model where parameters are perturbed by random noise, allowing the resampling step to provide Darwinian natural selection among the population of particles which favors parameter values consistent with the data.
The outer loop, $m=1,\dots,M$, iterates this BPF procedure while decreasing the magnitude of the perturbations, which is theoretically guaranteed to guide the parameters toward a neighborhood of the maximum likelihood estimate \citep{ionides15,ning23}.
Further discussion of BPF and IBPF is in Supplementary Sec.~{\suppSecIBPF}.
Using this maximization procedure, we constructed confidence intervals by profile likelihood, employing Monte Carlo adjusted profiles \cite{ionides17,ning21} to correct for Monte Carlo variability.

\vspace{3mm}

\noindent {\it Model criticism}.
A negative binomial autoregressive model was used to provide a non-mechanistic benchmark log-likelihood, as described in Supplementary Sec.~\suppSecBenchmark.
This model was also used to construct benchmark conditional log-likelihoods for each separate observation.
These, differenced from the corresponding SEAIR log-likelihoods, were used to define anomalies.
The anomalies were explored to identify data points which were poorly explained by the model (Supplementary Sec.~\suppSecResiduals).
In preliminary data analysis, these anomalies helped to identify some errors in the data which were subsequently corrected (Supplementary Sec.~\suppSecLiModel).

\vspace{3mm}

\noindent {\it Software environment}.
Numerical work was carried out in R \cite{R}. 
Models and data analysis methodology were developed in an R package, \code{metapoppkg}, which is additionally designed to assist reproducibility and extendability of our results.
Models in \code{metapoppkg} are implemented using \code{spatPomp} \cite{asfaw23arxiv} which provides a general representation of SpatPOMP models extending the POMP model representation in \code{pomp} \cite{king16}.

\section*{Data and code availability}
The \code{metapoppkg} R package, containing the data and software used for this article, is available at \url{https://github.com:jifanli/metapoppkg} and archived at \url{https://zenodo.org/records/10149233}.
The manuscript source code is available at \url{https://github.com:jifanli/metapop_article} and archived at \url{https://zenodo.org/records/10149258}.
This source code depends on the \code{metapoppkg} R package and other open-source software archived at \url{https://cran.r-project.org/}.

\section*{Acknowledgements}
This work was supported by National Science Foundation grants DMS-1761603 and DMS-1761612.
Portions of this research were conducted with Texas A\&M High Performance Research Computing and University of Michigan Advanced Research Computing.
We acknowledge discussions with Ethan Romero-Severson and Bryan Grenfell.

\arxiv{}{
\section*{Author contributions}
J.L., E.L.I. and N.N. designed the study and carried out the numerical analysis. J.L., E.L.I, A.A.K. and N.N. developed the software environment supporting the data analysis. All authors discussed and interpreted the results. E.L.I. drafted the manuscript, and all authors edited the manuscript.
}

\bibliographystyle{ieeetr}

%\bibliography{bib-metapop}

\end{document}

\date{\small Compiled \today,  using \proglang{R} 4.2.3, \pkg{metapoppkg} 0.1.18, \pkg{spatPomp} 0.34.1, and \pkg{pomp} 5.5.1.1.}
\title{Supplement to ``{\myTitle}''}
\author{\myAuthorList}
\maketitle

%\vspace{3mm}

\setcounter{tocdepth}{1}
\tableofcontents

 \newpage

\section{\secSep Specification of the models}
\label{sec:li23}

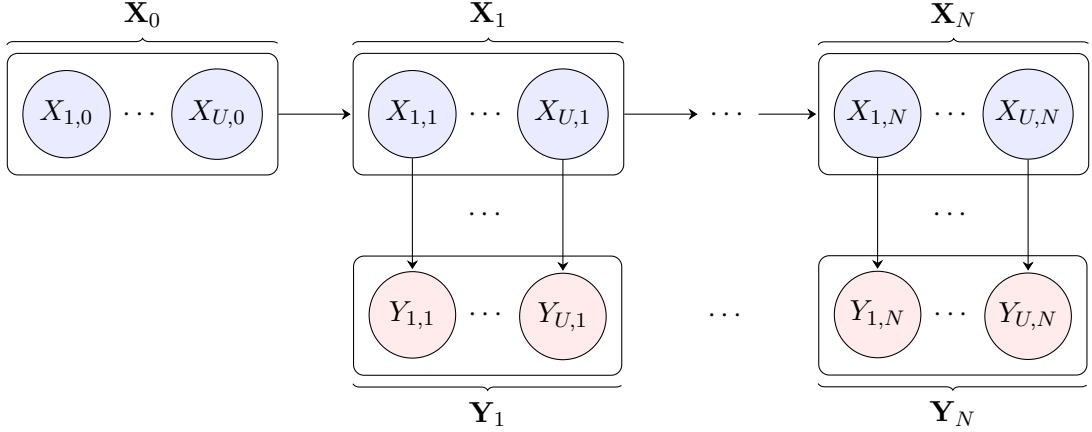
\begin{figure}[h]
  \centering
\begin{tikzpicture}
[
  state/.style = {draw, circle, fill=blue!8, minimum width = 1.15cm}, 
  obs/.style = {draw, circle, fill=red!8, minimum width = 1.15cm}, 
  >/.style={shorten >=0.25mm}, % redefine arrow to stop short of node
  >/.tip={Stealth[length=1.2mm,width=1.5mm]}, % redefine arrow style
  ssp/.style={draw, rounded corners,inner sep=2mm} % state space
]
\tikzset{>={}}; % this is needed to implement the arrow redefinition

		\node[state] (X10) {$X_{1,0}$}; 
		\node (D0) [right of=X10] {$\cdots$}; 
		\node[state] (XU0) [right of=D0] {$X_{U,0}$}; 
		\node[state] (X11) [right=1.4cm of XU0] {$X_{1,1}$}; 
		\node(XD1) [right of=X11] {$\cdots$}; 
		\node[state] (XU1) [right of=XD1] {$X_{U,1}$};
		\node[ssp, fit=(X10) (D0) (XU0)](X0){};
		\draw[decoration={brace,raise=3pt},decorate](X0.north west) -- node[above=6pt] {$\myvec{X}_0$} (X0.north east);
		\node[ssp, fit=(X11) (XD1) (XU1)](X1){};
		\draw[decoration={brace,raise=3pt},decorate](X1.north west) -- node[above=6pt] {$\myvec{X}_1$} (X1.north east);
		\draw[->] (X0) -- (X1);
		\node[obs] (Y11) [below=1.5cm of X11] {$Y_{1,1}$}; 
		\node (YD1) [right of=Y11] {$\cdots$}; 
		\node[obs] (YU1) [below=1.5cm of XU1] {$Y_{U,1}$};
		\node[ssp, fit=(Y11) (YD1) (YU1)](Y1){};
		\draw[decoration={brace,mirror,raise=3pt},decorate](Y1.south west) -- node[below=6pt] {$\myvec{Y}_1$} (Y1.south east);
		\draw[->] (X11.south) -- (Y11.north) node[midway, right=0.6cm]{$\cdots$};
		\draw[->] (XU1.south) -- (YU1.north);
    \draw [->] (X1.east) -- ++(\FixedLengthArrow) node[right] (midstates) {$\cdots$};
    \node[state,inner sep=2.5pt] (X1N) [right=1cm of midstates] {$X_{1,N}$}; 
		\node (XDN) [right of=X1N] {$\cdots$}; 
		\node[state,inner sep=2.5pt] (XUN) [right of=XDN] {$X_{U,N}$};
		\node[ssp, fit=(X1N) (XDN) (XUN)](XN){};
		\draw[decoration={brace,raise=3pt},decorate](XN.north west) -- node[above=6pt] {$\myvec{X}_N$} (XN.north east);
		\draw [->] (midstates) -- (XN);
		\node[obs,inner sep=2.8pt] (Y1N) [below=1.5cm of X1N] {$Y_{1,N}$}; 
		\node (YDN) [right of=Y1N] {$\cdots$}; 
		\node[obs,inner sep=2.8pt] (YUN) [below=1.5cm of XUN] {$Y_{U,N}$};
		\node[ssp, fit=(Y1N) (YDN) (YUN)](YN){};
		\draw[decoration={brace,mirror,raise=3pt},decorate](YN.south west) -- node[below=6pt] {$\myvec{Y}_N$} (YN.south east);
		\draw[->] (X1N.south) -- (Y1N.north) node[midway, right=0.6cm]{$\cdots$};
		\draw[->] (XUN.south) -- (YUN.north);
		\node (YD) [right=1cm of Y1] {$\cdots$};
	\end{tikzpicture}
\caption{
Diagram for a spatiotemporal partially observed Markov process (SpatPOMP) model,  adapted from \citep{asfaw23arxiv}.
The latent dynamic process is a continuous-time Markov chain taking value $\myvec{X}(t)=\big(X_0(t),\dots,X_U(t)\big)$ at time $t$.
At observation time $t_n$, the value of the latent process is denoted by $\myvec{X}(t_n)=\myvec{X}_{n}=\big(X_{1,n},\dots,X_{U,n}\big)$.
Noisy and/or incomplete observations of the latent process at this time are modeled by $\myvec{Y}_{n}=\big(Y_{1,n},\dots,Y_{U,n}\big)$.
} \label{fig:spatpomp}
\end{figure}

We give the mathematical description of the COVID-19 models described in Figure~{\MainFigModelDiagram} and Table~{\MainResultsTable} of the main text.
These metapopulation models are partially observed Markov process (POMP) models with additional spatial structure, known as SpatPOMP models \citep{asfaw23arxiv}.
The diagram in Figure~\ref{fig:spatpomp} defines a general SpatPOMP corresponding to a metapopulation with $U$ spatial units.
The latent dynamic state is $\myvec{X}(t)=\big(X_1(t),\dots,X_U(t)\big)$, which we also write as $\myvec{X}(t)=X_{1:U}(t)$.
The observation model at time $t_n$ is $\myvec{Y}_{n}=\big(Y_{1,n},\dots,Y_{U,n}\big)=Y_{1:U,n}$, for $n$ taking values in $\seq{1}{N}$.
The data, $\myvec{y}^*_n=y^*_{1:U,n}$, are modeled as a realization of the random variable $Y_{1:U,n}$.

For each city, $\unit$ in $\seq{1}{\Unit}$, with $\Unit=373$, we model the state at time $t$ as
\begin{equation}
X_{\unit}(t)=\big(S_{\unit}(t),E_{\unit}(t), A_{\unit}(t), I_{\unit}(t), R_{\unit}(t),
  C^a_{\unit}{t}, C^b_{\unit}(t), C_{\unit}(t)\big),
\end{equation}
where each individual in the city is in exactly one of the compartments: susceptible ($S_{\unit}$), exposed ($E_{\unit}$), infected and infectious but asymptomatic ($A_{\unit}$), infected and infectious and symptomatic ($I_{\unit}$), and recovered or removed ($R_{\unit}$).
The additional case reporting compartments, $C^a_{\unit}$, $C^b_{\unit}$ and $C_{\unit}$ are used to describe reporting delay.
Individuals entering $I_{\unit}$ are simultaneously added to $C^a_{\unit}$, from which they transition to $C^b_{\unit}$ and subsequently to the observable compartment, $C_{\unit}$.
For notational convenience, we introduce a transport compartment, $T$, which accounts for all individuals traveling between cities.
The complete collection of compartments is therefore
\begin{equation}
{\mathbb C} = \big\{ S_{1:\Unit}, E_{1:\Unit},A_{1:\Unit},I_{1:\Unit},R_{1:\Unit},C^a_{1:\Unit},C^b_{1:\Unit},C_{1:\Unit},T\}.
\nonumber
\end{equation}
We let $N_{VW}(t)$ count the directional transitions between $V$ to $W$ for any pair of compartments in ${\mathbb C}$, and we write $dN_{VW}$ for an infinitesimal increment, $N_{VW}(t+dt)-N_{VW}(t)$.
We can write $\mathbf{X}(t)=\big(X_1(t),\dots,X_\Unit(t)\big)$ in terms of its value $\mathbf{X}(t_0)$ at an initial time $t_0$ together with the flow equations:
\begin{eqnarray}
\label{eq:dS}
dS_{\unit} &=& -dN_{S_{\unit}E_{\unit}} + dN_{TS_{\unit}} -  dN_{S_{\unit}T},
\\
dE_{\unit} &=& dN_{S_{\unit}E_{\unit}} - dN_{E_{\unit}A_{\unit}} - dN_{E_{\unit}I_{\unit}} + dN_{TE_{\unit}} -  dN_{E_{\unit}T},
\\
dA_{\unit} &=& dN_{E_{\unit}A_{\unit}} + dN_{TA_{\unit}} -  dN_{A_{\unit}T} - dN_{A_{\unit}R_{\unit}},
\\
dI_{\unit} &=& dN_{E_{\unit}I_{\unit}} - dN_{I_{\unit}R_{\unit}},
\\
dR_{\unit} &=& dN_{I_{\unit}R_{\unit}} + dN_{A_{\unit}R_{\unit}},
\\
dC^a_{\unit} &=& dN_{E_{\unit}I_{\unit}} - dN_{C^a_{\unit}C^b_{\unit}},
\\
dC^b_{\unit} &=& dN_{C^a_{\unit}C^b_{\unit}} - dN_{C^b_{\unit}C_{\unit}},
\\
\label{eq:dC}
dC_{\unit} &=& dN_{C^b_{\unit}C_{\unit}}.
\end{eqnarray}
In this model, individuals travel between cities only when in compartments S (susceptible), A (infected and infectious but unreported, generally asymptomatic or mildly symptomatic),  and E (exposed with a latent infection).
Also, note that we do not match up each individual entering and leaving $T$, so there can be small stochastic variation in the total population.

Each transition $dN_{VW}$ has an associated rate, $\mu^{}_{VW}$, which may depend on the state of other compartments, or on covariate processes, on parameters, or on time.
For all compartments other than the source/sink compartment, $T$, it is convenient to specify rates per capita.
For transitions which enter a compartment from $T$, we specify a total rate.
The non-zero transition rates are therefore as follows:
\begin{eqnarray}
\label{eq:muSE}
\rate_{S_{\unit}E_{\unit}} &=& \beta \left(\frac{I_{\unit}(t)+ \relativeTransmission A_{\unit}(t)}{\pop_{\unit}(t)}\right)
d\Gamma_{\unit}\big/dt,
\\
\rate_{S_{\unit}T} \eqsep = \eqsep \rate_{E_{\unit}T} \eqsep = \eqsep \rate_{A_{\unit}T} &=& \mobilityFactor \sum_j \frac{M_{\unit j}(t)}{\pop_{\unit}-I_{\unit}},
\\
\rate_{TS_{\unit}} &=& \mobilityFactor \sum_j \frac{M_{j \unit}(t)\, S_{j}}{\pop_{j}-I_{j}},
\\
\rate_{TE_{\unit}} &=& \mobilityFactor \sum_j \frac{M_{j \unit}(t)\, E_{j}}{\pop_{j}-I_{j}},
\\
\rate_{TA_{\unit}} &=&\mobilityFactor \sum_j \frac{M_{j \unit}(t)\,  A_{j}}{\pop_{j}-I_{j}},
\\
 \eqsep \rate_{E_{\unit}I_{\unit}}  &=&  \alpha/Z,
 \\
 \rate_{E_{\unit}A_{\unit}} \eqsep &=& (1-\alpha)\big/ Z
\\
\rate_{A_{\unit}R_{\unit}} \eqsep = \eqsep \rate_{I_{\unit}R_{\unit}}  &=& 1/D,
\\
\rate_{C^a_{\unit}C^b_{\unit}} \eqsep = \eqsep \rate_{C^b_{\unit}C_{\unit}}  &=& 2\big/\diagnosisDelay.
\label{eq:muCbC}
\end{eqnarray}
We define $d\Gamma_{\unit}/dt$ in (\ref{eq:muSE}) as non-negative multiplicative gamma white noise with variance parameter $\sigma_{SE,\unit}$.
That is, $\Gamma_{\unit}(t)$ is a gamma process with stationary independent increments, such that $\Gamma_{\unit}(t)-\Gamma_{\unit}(s)$ is gamma distributed with mean $t-s$ and variance $\sigma_{SE}(t-s)$.
Equations (\ref{eq:dS})-(\ref{eq:muCbC}) therefore specify an overdispersed continuous time Markov process via the limit of a discrete time Euler approximation as the discretization step approaches zero \citep{breto09,breto11}.

The time-varying transport matrix, $M_{\unit j}(t)$, describes the rate of individuals moving from city $u$ to city $j$ at time $t$.
It is modeled as piecewise constant for each day, and its construction is detailed in Section~\ref{sec:mobility}.
To compensate for imperfect transport data, we include a calibration constant, $\mobilityFactor$.
The model parameters are described in Table~\ref{tab:parameters}, together with their units and their fitted values.

Data for city $\unit$ at time $t_n$ is an official report $y^*_{\unit,n}$ recording new cases since time $t_{n-1}$.
The data are modeled as a realization of a random variable $Y_{\unit,n}$ which measures $C_{\unit}(t_n)-C_{\unit}(t_{n-1})$.
The measurement model asserts that $Y_{\unit,n}$ is a discretized normal random variable with mean
\begin{equation}
\label{eq:measure1}
C_{\unit,n} = C_{\unit}(t_{n})-C_{\unit}(t_{n-1}),
\end{equation}
and variance
\begin{equation}
\label{eq:vunit_measure}
V_{\unit,n}=C_{\unit,n}+ \tau^2C_{\unit,n}^2,
\end{equation}
including both Poisson scale variability and the possibility of overdispersion.
Thus,
\begin{equation}\nonumber
\prob\big(Y_{u,n}=y_{u,n}\given C_{u,n}\big)=\Phi\big(y_{u,n}+0.5 \giventh C_{u,n},V_{u,n}\big) - \Phi\big(y_{u,n}-0.5 \giventh C_{u,n},V_{u,n}\big),
\end{equation}
where $\Phi$ is the normal cumulative distribution function.
If $y_{u,n}=0$, we replace $ \Phi\big(y_{u,n}-0.5 \giventh C_{u,n},V_{u,n}\big)$ by $\Phi\big(-\infty \giventh C_{u,n},V_{u,n}\big)=0$.

Our models {\LiMobility}, {\LiParams}, {\RevisedModelUnconstrained} and {\RevisedModelConstrained} are extensions of the model of \citet{li20}.
The original model of \citet{li20}, which we call \code{li20}, represents the dynamic model by a system of ordinary differential equations with random rates, as discussed further in Section~\ref{sec:li20}.
The general model including {\LiMobility}, {\LiParams}, {\RevisedModelUnconstrained} and {\RevisedModelConstrained}, which we call \code{li23}, represents the models as continuous-time Markov chains.
In addition to this change,  \code{li23} considers two model aspects not investigated by \citet{li20}: (i) overdispersed process noise; (ii) an adjusted movement matrix to ensure that all cities are connected.
In  {\LiMobility}, these two features are turned off, and so this model is very similar to \code{li20}, as documented in Section \ref{sec:li20}, below, which discusses \code{li20} in more detail.
{\LiMobility} and {\LiParams} have the constraint that $\sigma_{SE}=0$.
{\LiMobility} has an additional constraint that $\tau=0$ since this parameter was not included in the dynamic model of \cite{li20}.
Another difference between these models is that {\LiMobility} uses the mobility data from \citep{li20} whereas {\LiParams} (together with {\RevisedModelUnconstrained} and {\RevisedModelConstrained}) use the modification described in Section~\ref{sec:mobility}.
{\RevisedModelUnconstrained} involves estimation of all the parameters estimated by \cite{li20}, with the addition of $\sigma_{SE}$ and $\tau$.
{\RevisedModelConstrained} adds the additional constraints that $\latency^{\after}=\latency^{\before}$  and $\infectiousPeriod^{\after}=\infectiousPeriod^{\before}$.

All the model parameters were specified to be shared between units.
The model can be extended to define distinct unit-specific values for each parameter, and in some situations this is helpful \citep{ionides22,whitehouse23}.
We developed an R package \pkg{metapoppkg} to provide a data analysis environment for our numerical work, described further in Section~\ref{sec:metapoppkg}.
The \code{li23} and \code{li20} models are implemented by the \code{li23()} and \code{li20()} constructor functions in \pkg{metapoppkg}.
The resulting model objects have class \class{spatPomp} which provides access to the inference and visualization tools in the R package \pkg{spatPomp} \citep{asfaw23arxiv} as well as \pkg{pomp} \citep{king16}.
Here, we focus on likelihood evaluation via the block particle filter, implemented as \code{bpfilter}, and parameter estimation via the iterated block particle filter, \code{ibpf} (discussed in Section~\ref{sec:bpfilter}).
The ensemble Kalman filter and iterated ensemble Kalman filter, as employed by \citep{li20}, are implemented as \code{enkf} and \code{ienkf}, respectively,  and are discussed further in Section~\ref{sec:enkf}.

%%%% parameter table: search for pppppppppppp %%%%%%%%%%%

\begin{table}
\begin{tabular}{crc|rc|rc|lc} & 
  {\LiMobility} & CI &
  {\RevisedModelUnconstrained} & CI &
  {\RevisedModelConstrained} & CI&
  interpretation \& units 
\\
\hline
$\mobilityFactor$ &
  1.36 &(1.27,1.45)&
  2.34 &
    (2.07,2.54)&  
  2.87 &
    (2.67,3.19)&  
  Mobility factor 
  \\  
$\tau$ &
  0 & fixed &
  0.28 &
    (0.27,0.33)&    
  0.32 &
    (0.29,0.35)&  
  Measurement noise 
  \\
$\sigma_{SE}$ &
  0 & fixed &
  2.08 &
    (1.93,2.27)&    
  1.77 &  
    (1.58,1.92)&  
  Dynamic noise (day$^{1/2}$)
  \\
$E_0$ &
  0--2000 & NA &
  2712 &
    (2381,3122)&    
  3477 &  
    (2730,3846)&  
  \EzeroDescription 
  \\
$A_0$ &
  0-2000 &NA&
  0 &
    (0,307)&    
  0 &  
    (0,440)&  
  \AzeroDescription 
  \\
\hline  
$\beta^{\before}$ &
  1.12 &(1.06,1.09)&
  0.73 &
    (0.70,0.78)&    
  0.97 &
    (0.93,1.05)&  
  Transmission rate (day$^{-1}$)
  \\
$\relativeTransmission^{\before}$ &
  0.55 &(0.46,0.62)&
  1.00 &
    (0.96,1.00)&    
  1.00 &
    (0.93,1.00)&  
  Relative transmission 
  $\mu$
  \\
$\latency^{\before}$ &
  3.69 &(3.30,3.96)&
  0.55 &
    (0.28,0.83)&  
  $^\ast$0.72 &
    (0.48,0.97)&
  Latent period (day)
  \\
$\infectiousPeriod^{\before}$ &
  3.47 &(3.15,3.73)&
  35.0 &
    (5.0,35.0)&  
   $^\dagger$3.87 &
    (3.69,4.10)&
  Infectious period (day)
  \\
$\reportRate^{\before}$ &
  0.14 &(0.10,0.18)&
  0.11 &
    (0.09,0.12)&  
  0.08 &
    (0.07,0.08)&
  \alphaDescription 
  \\
$\Rzero^{\before}$ &     
  2.38 &(2.03,2.77)&
  13.07 &
    (11.96,13.72)&  
  3.51 &
    (3.31,3.72)&
  Basic reproductive number 
  \\
$\diagnosisDelay^{\before}$ &
  9.00 &fixed&
  9.00 &fixed&
  9.00 &fixed&
  Diagnosis delay (day)
  \\
\hline
$\beta^{\after}$ &
  0.35 &(0.28,0.45)&
  0.24 &
    (0.21,0.26)&  
  0.22 &
    (0.21,0.25)&
  Transmission rate (day$^{-1}$)
  \\
$\relativeTransmission^{\after}$ &
  0.43 &(0.31,0.61)&
  0.61 &
    (0.52,0.82)&     
  0.78 &
    (0.66,0.90)&   
  Relative transmission 
  \\
$\latency^{\after}$ &
  3.42 &(3.30,3.65)&
  4.23 &
    (3.39,5.04)&     
   $^\ast$0.72 &
    (0.48,0.97)&   
  Latent period (day)
  \\
$\infectiousPeriod^{\after}$ &
  3.31 &(2.96,3.88)&
  2.36 &
    (2.16,2.51)&     
  $^\dagger$3.87 &
    (3.69,4.10)&   
  Infectious period (day)
  \\
$\reportRate^{\after}$ &
  0.69 &(0.65,0.72)&
  0.38 &
    (0.34,0.47)&     
  0.48 &
    (0.39,0.52)&   
  \alphaDescription 
  \\
$\Rzero^{\after}$ &    
  0.95 &(0.83,1.16)&
  0.51 &
    (0.49,0.58)&     
  0.70 &
    (0.65,0.77)&   
  Basic reproductive number 
  \\
$\diagnosisDelay^{\after}$ &
  6.00 &fixed&
  6.00 &fixed&
  6.00 &fixed&
  Diagnosis delay (day)
  \\
\hline
\end{tabular}
\caption{Parameter estimates and their confidence intervals (CIs).
The parameter estimates and confidence intervals for {\LiMobility} come from \cite{li20}.
The values for {\RevisedModelUnconstrained} and {\RevisedModelConstrained} come from profile likelihood plots shown in Sections~\ref{sec:unconstrained_estimation} and~\ref{sec:constrained_estimation} respectively.
Top block of rows: parameters constant through time.
Middle block: parameters estimated for Jan 10-Jan 23.
Bottom block: parameters estimated for Jan 24-Feb 8.
Parameters without specified units are dimensionless.
For {\RevisedModelConstrained}, $\latency^{\before}=\latency^{\after}$, so the two values marked by $^\ast$ are constrained to be equal; $^\dagger$ denotes the other constraint $\infectiousPeriod^{\before}=\infectiousPeriod^{\after}$.
We calculated $\Rzero^{\before}=\big(\alpha^{\before}+(1-\alpha^{\before})\mu^{\before}\big)\infectiousPeriod^{\before}\beta^{\before}$ and  $\Rzero^{\after}=\big(\alpha^{\after}+(1-\alpha^{\after})\mu^{\after}\big)\infectiousPeriod^{\after}\beta^{\after}$.
}\label{tab:parameters}
\end{table}

\begin{knitrout}
\definecolor{shadecolor}{rgb}{0.969, 0.969, 0.969}\color{fgcolor}\begin{figure}

{\centering \includegraphics[width=6.5in]{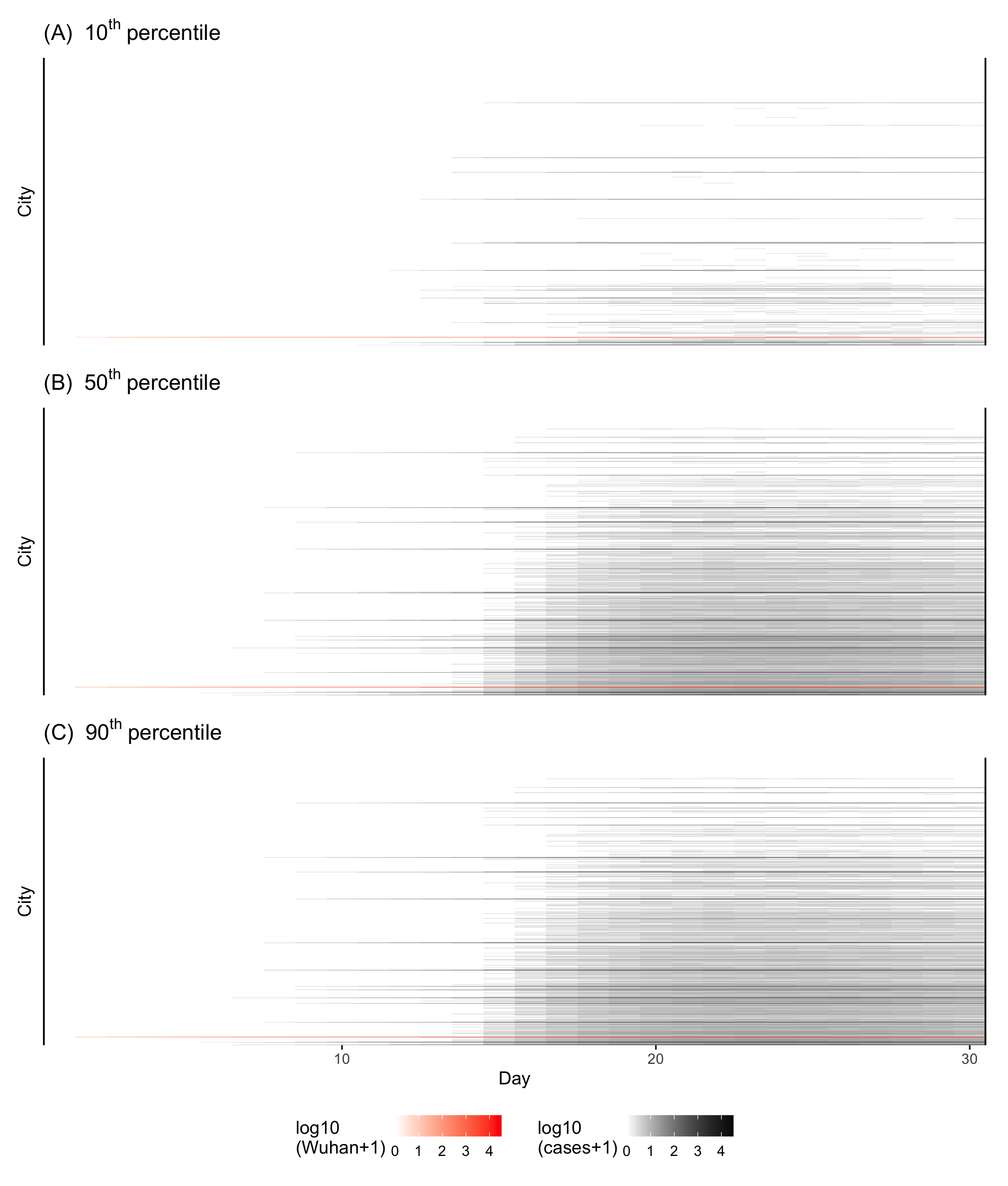} 

}

\caption[Simulated daily case reports for model {\RevisedModelUnconstrained}, showing the 10th, 50th and 90th percentiles]{Simulated daily case reports for model {\RevisedModelUnconstrained}, showing the 10th, 50th and 90th percentiles. Within each panel, cities are ordered by population, largest on the bottom row.}\label{fig:percentile_plot}
\end{figure}

\end{knitrout}

Table~\ref{tab:parameters} shows some substantial differences between our parameter estimates and those of \citet{li20}.
Model~{\RevisedModelUnconstrained}, which has considerably higher likelihood than {\LiMobility} due to the inclusion of dynamic process noise, obtains its highest likelihoods when the pre-lockdown infectious period parameter is unrealistically large.
In this model, a long duration of infection pre-lockdown does not cause problems because individuals leave the infected class quickly once lockdown arrives.
Constraining the durations of latency and infection to be the same before and after lockdown changes this, without having substantial effects on other parameter estimates.
This occurs at the expense of
$-9088.2 - (-9116.5) = 28.3$
units of log-likelihood.
We cannot readily see why the data prefer a mechanistically implausible infectious period pre-lockdown.
However, weak identifiability is not surprising in a complex model with many parameters that is required to fit fairly sparse amounts of data pre-lockdown.
Some model misspecification is also inevitable for a mathematical model of a biological system.
When weak identifiability and model misspecification co-occur, one possible result is scientifically implausible parameter estimates.

The data are compared to simulations from models {\LiMobility} and {\RevisedModelConstrained} in Figure~{\MainFigSimulations} of the main text.
The variability in {\RevisedModelUnconstrained} and {\RevisedModelConstrained} is explicitly included in the model and fitted to the data; it therefore matches the variability in the data more closely than  {\LiMobility}.
If many simulations are made from {\RevisedModelUnconstrained}, the pointwise 10$^{\mathrm{th}}$ percentile is similar to a simulation from {\LiMobility} (Figure~\ref{fig:percentile_plot}).
The similarity between model {\LiMobility} and \code{li20} is evident by comparing Figure~{\MainFigSimulations}(B) with Figure~\ref{fig:li20_plot} in Section~\ref{sec:li20}.
The code and parameters for the simulation from \code{li20} are taken directly from \citet{li20}.

\subsection{\secSep Reporting Delay}

The \code{li23} and \code{li20} models describe the reporting process via the case report compartments, $C^a_u$, $C^b_u$ and $C_u$ for each city, $u$.
When an individual transitions from $E_u$ to the reportable infectious state, $I_u$, an individual is also added to the start of the reporting process by incrementing $C^a_u$.
The counts in compartments $C^a_u$, $C^b_u$ and $C_u$ do not affect the transmission dynamics; they only part of the measurement model.
For this reason, they are denoted by octagons rather than squares in the diagrammatic representation (Figure~1, main text).

\citet{li20} described transitions from $C^a_u$ to $C_u$ using a gamma delay model, with each individual arriving in $C^a_u$ transitioning to $C_u$ after a random delay distributed as $G(a,T_d/a)$, the gamma distribution with mean $T_d$ and variance $T_d^2/a$.
Based on analysis of early confirmed cases, and preliminary exploration of the model, they specified $a=1.85$, $T_d=9$ before January 23, and $T_d=6$ after January 23.
We use their values of $T_d$ but use $a=2$ in order to obtain a Markovian representation, where the gamma delay is represented as the sum of two exponential delays, formalized using a compartment $C^b_u$ intermediate between $C^a_u$ and $C_u$. 

Our interpretation of reporting delay leads to a small discrepancy between our \code{li20} model and the model actually specified by \citet{li20}.
However, the discrepancy is small.
Further, the Markovian property is necessary for inference using either the ensemble Kalman filter or block particle filter.
Thus, this discrepancy closes a small gap between the model specified by \citet{li20} and the methods which they (and we) use to analyze the model.

\subsection{\secSep Mobility Data} \label{sec:mobility}

\begin{knitrout}
\definecolor{shadecolor}{rgb}{0.969, 0.969, 0.969}\color{fgcolor}\begin{figure}

{\centering \includegraphics[width=3.5in]{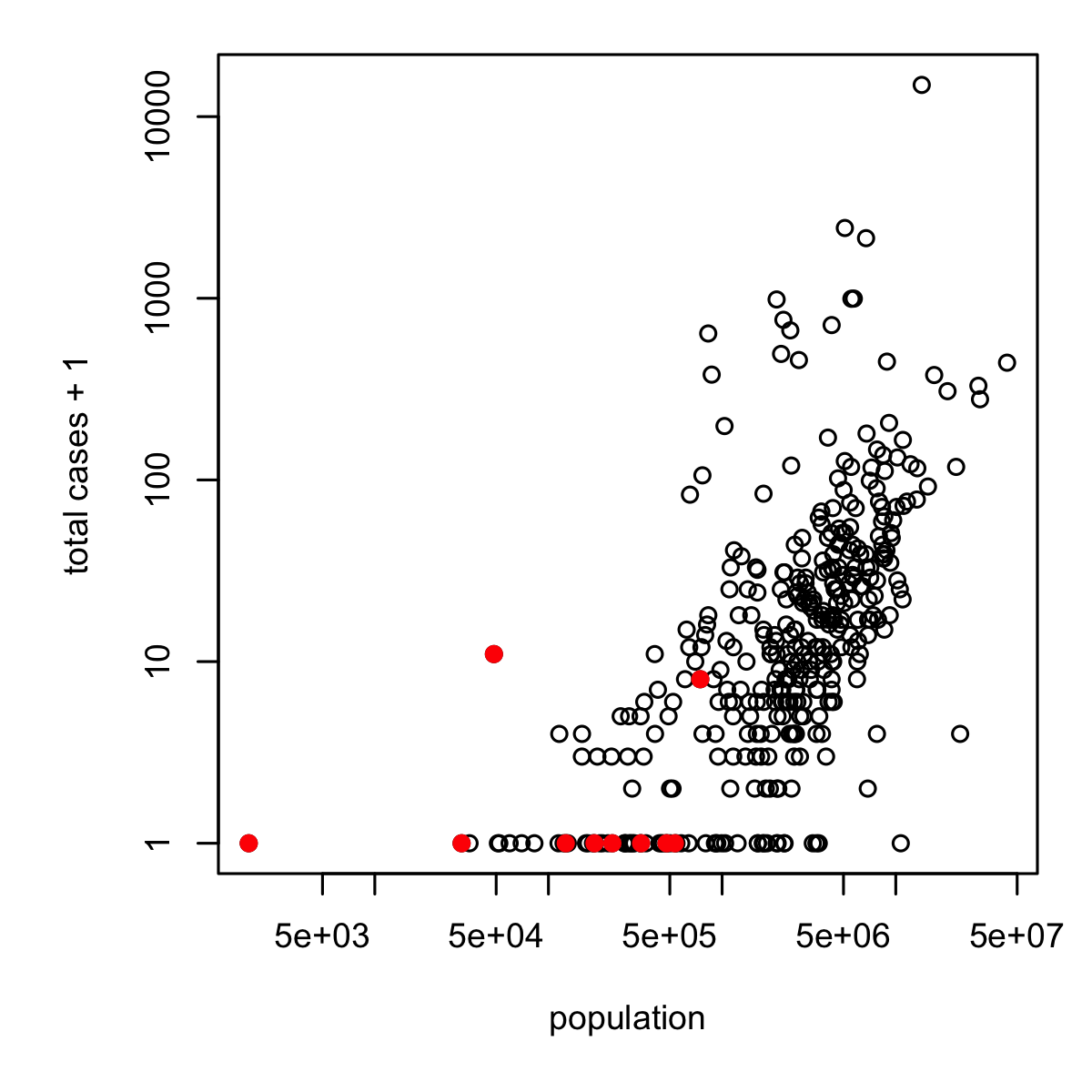} 

}

\caption[Total cases, January 10 to February 8, for each city, plotted against mean population size]{Total cases, January 10 to February 8, for each city, plotted against mean population size.  Cities with no arriving travelers recorded in the mobility data are shown as solid red points.}\label{fig:zeros_plot}
\end{figure}

\end{knitrout}

Figure~\ref{fig:zeros_plot} shows the 10 cities which have no incoming travelers in the mobility dataset compared to other cities.
We see that the cities modeled as having no sources by \citet{li20} did have relatively few reported cases for their city size, but not a complete absence of cases.

To capture individual movement among the 373 cities simulated in the metapopulation model, \citet{li20} used human mobility data from the Tencent location-based service used in popular Tencent mobile phone applications, such as Wechat, QQ, and Baidu Maps.
High resolution Tencent data were available for 2018, so they assumed the travel patterns captured in 2018 during the New Year celebrations (Chunyun) are similar to those of the analogous time period during 2020, prior to January 23 travel restrictions.
In total, 92,248 inter-city travel records were used to represent travel during January 10-23.
In the Tencent mobility data, for each day, the top 10 outflows from each of 373 Chinese cities were recorded.
For city-to-city connections for which only some of the days in this two-week time period rank in the top 10, \citet{li20} linearly interpolated missing daily outflow values. 

This procedure resulted in reasonable mobility estimates for most cities, but some cities remained disconnected, with no estimated incoming travelers (Figure~\ref{fig:mobility_travel_vs_size}, A and B).
Several small cities with few cases might be expected not to have a large impact on the overall analysis.
However, if their case reports have likelihood $0$ under a model then they can lead to a log-likelihood of $-\infty$ even for an otherwise suitable model.

We therefore added a small amount of additional movement between cities based on a gravity model,
\begin{equation}
M_{uj}(t) = M^{\mathrm {li20}}_{uj}(t) + \frac{{\mathcal F} \, \bar{d}}{\bar P(t_0)}\times \frac{ P_u(t_0)  \, P_j(t_0)}{d_{uj}},
\end{equation}
where $M^{\mathrm {li20}}_{uj}(t)$ is the movement rate from city $u$ to $j$ at time $t$ used by \citet{li20}, $P_u(t_0)$ is the initial population in city $u$; $\bar P(t_0)$ is the average population across all 373 cities; $d_{uj}$ is the distance from city $u$ to city $j$; $\bar d$ is the average of this distance over all $(373\times 372)/2$ pairs; $\mathcal F$ is a mobility correction factor which we took as ${\mathcal F}=20$ based on assessment of diagnostic plots.
Figure~\ref{fig:mobility-graph-plot} shows that this modification does not provide a major distortion to the pattern of travel from the movement data.
This figure displays only day 1 (January 10), but other days show similar patterns.
Figure~\ref{fig:mobility_travel_vs_size} gives further evidence for this; the modification is sufficient to move the zero travel records toward the main body of data, but not enough to result in other qualitative changes.

\begin{knitrout}
\definecolor{shadecolor}{rgb}{0.969, 0.969, 0.969}\color{fgcolor}\begin{figure}

{\centering \includegraphics[width=\maxwidth]{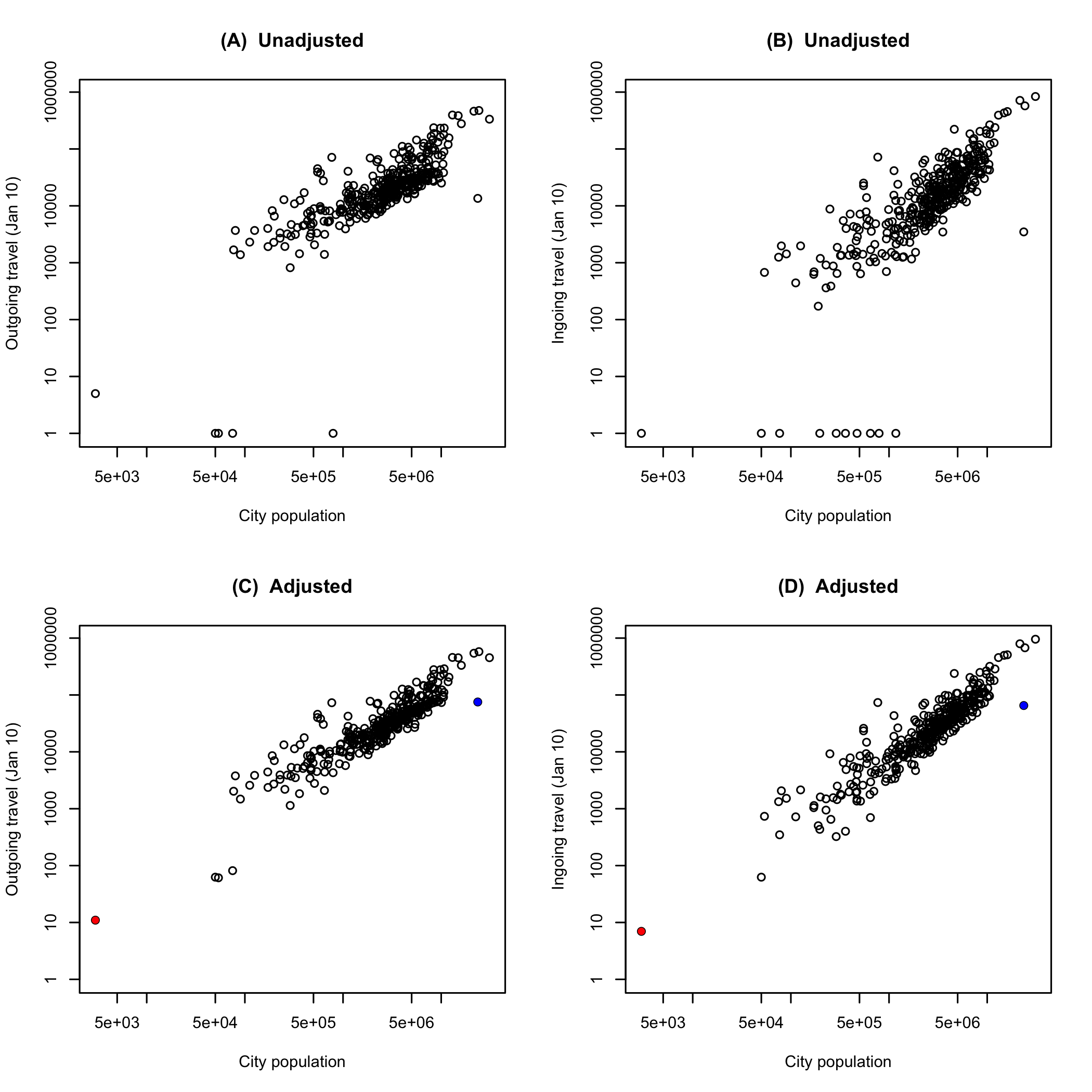} 

}

\caption[Total ingoing and outgoing travel plotted against city size]{Total ingoing and outgoing travel plotted against city size: (A,B) without an adjustment to ensure connectivity; (C,D) with the adjustment. The remaining outliers after adjustment are Sansha (red) and Taiwan (blue)}\label{fig:mobility_travel_vs_size}
\end{figure}

\end{knitrout}

\begin{knitrout}
\definecolor{shadecolor}{rgb}{0.969, 0.969, 0.969}\color{fgcolor}\begin{figure}

{\centering \includegraphics[width=6in]{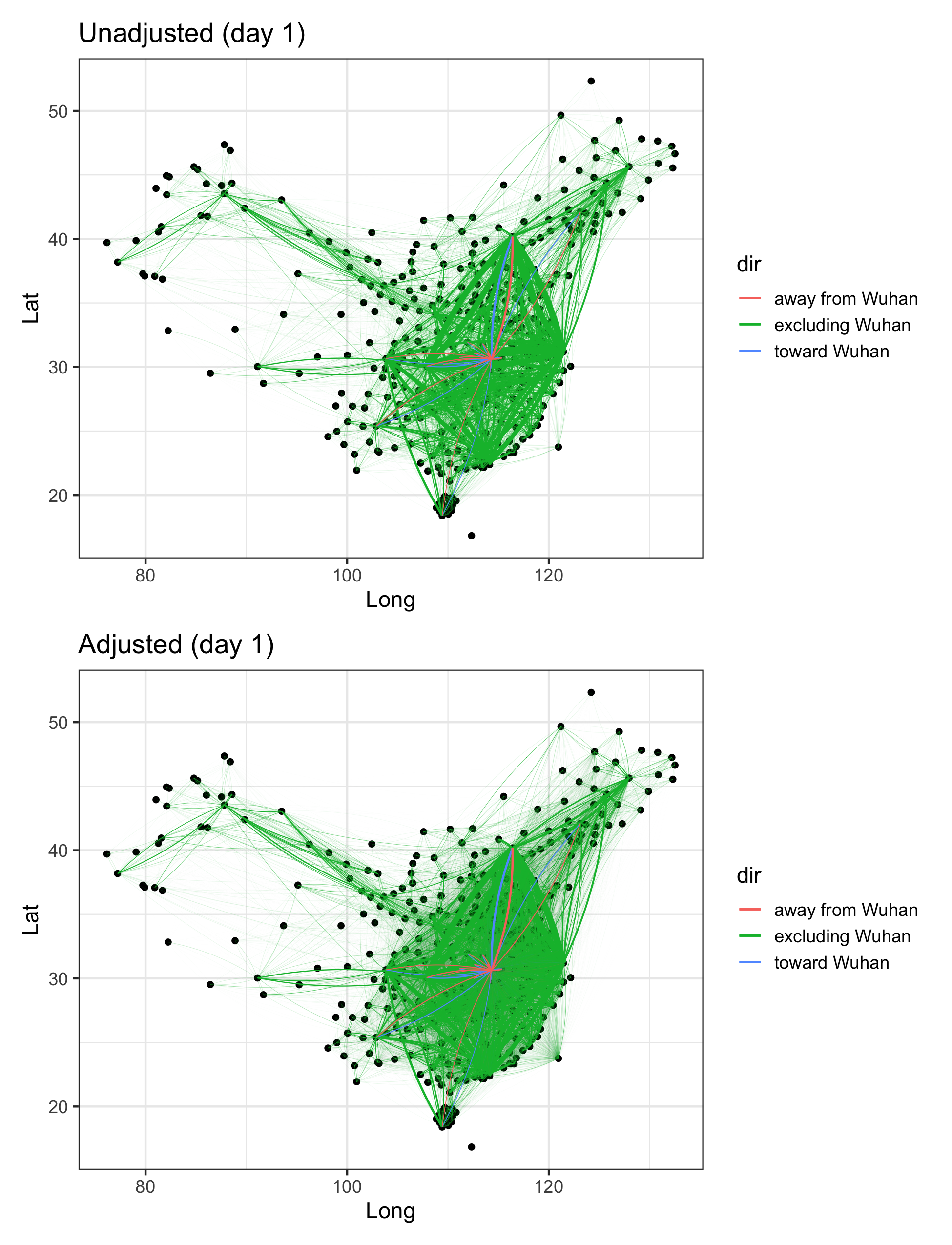} 

}

\caption[Mobility graph for day 1]{Mobility graph for day 1. Top: without adjustment to ensure full connectivity. Bottom: with adjustment. Edge thickness is proportional to movement.}\label{fig:mobility-graph-plot}
\end{figure}

\end{knitrout}

%\clearpage

\subsection{\secSep Comparison with the model and data of \citet{li20}}
\label{sec:li20}

\begin{knitrout}
\definecolor{shadecolor}{rgb}{0.969, 0.969, 0.969}\color{fgcolor}\begin{figure}

{\centering \includegraphics[width=6.5in]{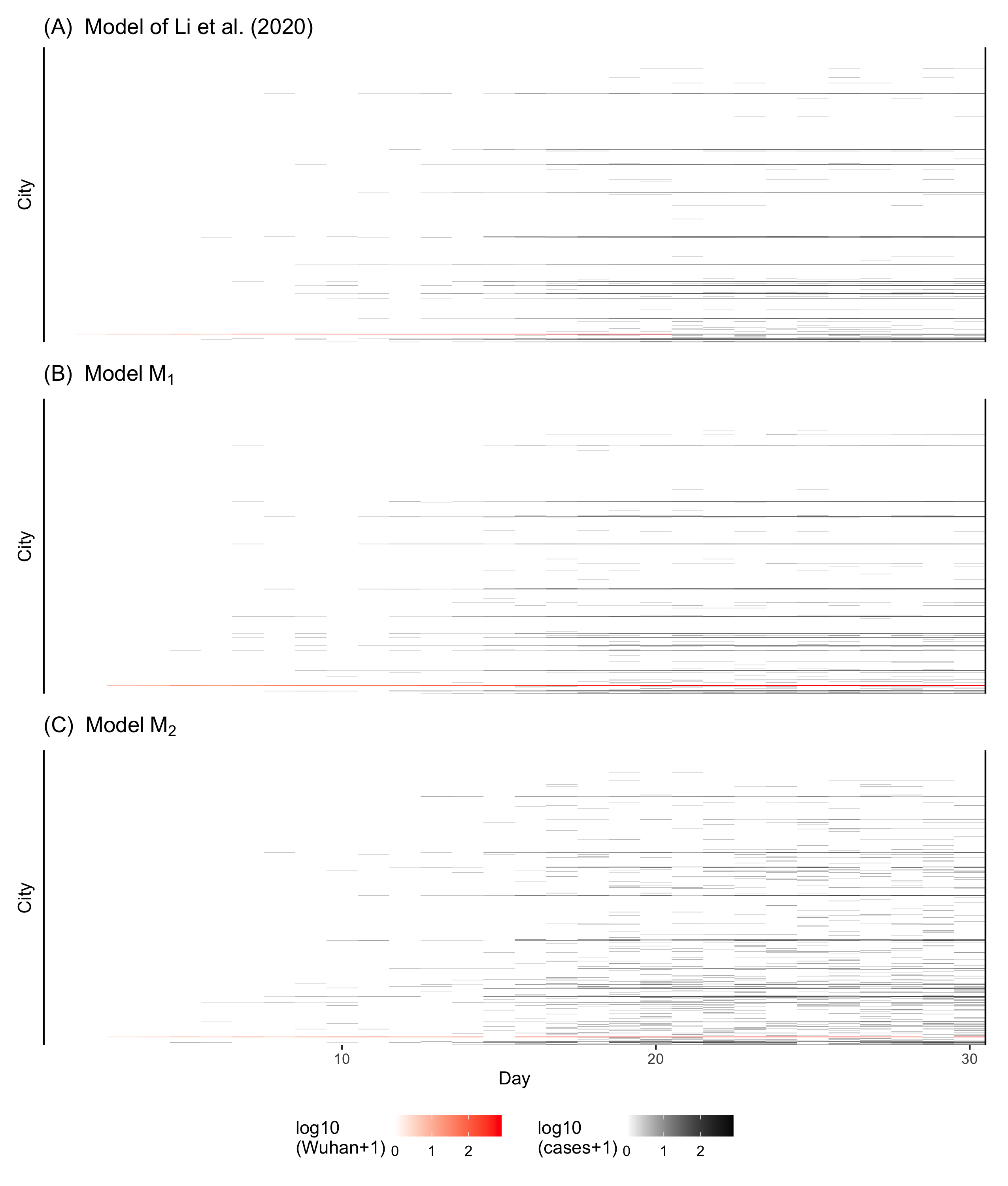} 

}

\caption{Simulations for 373 cities, comparing (A) a simulation from the model code and parameters provided by \citet{li20}; (B) a simulation from model {\LiMobility}; (C) a simulation from model~{\LiParams}. Within each panel, cities are ordered by population, largest on the bottom row.}\label{fig:li20_plot}
\end{figure}

\end{knitrout}

\citet{li20} specified the compartment model as a set of ordinary differential equations, with Poisson noise on rates, solved using a 4th-order Runge-Kutta scheme.
This approach constructs a discrete-time SpatPOMP, with the time discretization corresponding to the measurement times.
The continuous-time SpatPOMP representation of \code{li23} has some practical advantages:
\begin{enumerate}
\item It leads to simpler code. For example, compare the representation of our dynamic model in the R package \pkg{metapoppkg} function \code{li23} with either the code provided by \citet{li20} or our direct adaptation of this code in the function \code{li20} in our \texttt{metapoppkg} package.
Code simplicity facilitates debugging and consideration of model variations.
\item It allows inclusion of white noise on transmission rates to generate dynamic overdispersion \citep{breto09,he10,breto11,stocks20}.
This can lead to better fits to data (measured by likelihood) as well as avoiding over-confident predictions resulting from models that cannot adequately explain the variability in the data.
\end{enumerate}
Potentially, similar principles could be incorporated into \code{li20}, but the SpatPOMP framework makes the generalization more straightforward.

We implemented the \code{li20} model of \citet{li20} within the framework of the R package \pkg{spatPomp} \citep{asfaw23arxiv}.
The model is constructed by the \pkg{metapoppkg} function \code{li20()}.
This permits reproduction of the methodology used by \citet{li20}  via the \pkg{spatPom} function \code{ienkf}.
Our implementation of \code{li20} incorporates code adapted from supplementary information provided by \citet{li20}.
Simulation from \code{li20} using the parameters of \citet{li20} is therefore essentially equivalent to the simulation used by \citet{li20}.
In Figure~\ref{fig:li20_plot}, we compare a simulation from the code of \citet{li20} with simulations from models~{\LiMobility} and~{\LiParams}.
We see that \code{li20} and {\LiMobility} look similar; comparing with Figure~{\MainFigSimulations} in the main text, we see that both these models have distinctly less variability than the data.
The additional variability in {\LiParams} makes it look superficially more like the data, but the additional stochasticity is all ascribed to reporting variability in {\LiParams}; dynamic noise allows a better fit to the data, as documented formally in Table~{\MainResultsTable} and apparent from Figure~{\MainFigSimulations}.

The dataset analyzed by \citet{li20} included 375 cities, but we study only 373.
We found that two of the 375 cities are duplicates, with two slightly different names for the same town having the same location.
Also, the island of Hainan was included together with its separate counties---Hainan has a different administrative structure from other Chinese provinces, and does not have prefecture cities.
We removed the aggregated region of Hainan.
We modified some of the population values used by \citet{li20}.
When there was a major discrepancy between the value in their dataset and the prefecture population reported by Wikipedia, we took the latter.
The largest change was updating the population of Ezhou to 1,079,353 from 59,500.
Complete details of all our modifications to the dataset used by \citet{li20} are reported in the \code{metapop} package.

\section{\secSep Benchmark statistical models}

Basic statistical models, such as linear regression models or autoregressive-moving average (ARMA) time series models, or even independent random sample models, provide a baseline estimate of the predictability of the system under investigation.
A standard measure of this predictability is the log-likelihood \citep{gneiting07}, and it is therefore appropriate to compare log-likelihoods for different models calculated for the same data.
A sophisticated mechanistic model might be expected to have higher predictive skill, and therefore a higher log-likelihood, than a simple statistical model.
However, mechanistic models may be informative for what they cannot explain as well as what they can:
so far as a mechanistic model captures current understanding of the science of a system, we are interested to know when and where this science is inadequate to explain the data.
By contrast, statistical models are designed solely to provide a statistical fit.
If a mechanistic model fits substantially worse than a simple statistical model, one may infer that there is considerable room to improve the mechanistic model.
If the mechanistic model is competitive with non-mechanistic alternatives, regardless of whether its likelihood is actually higher, we infer that the mechanistic model provides a plausible explanation of the data.
Plausible mechanistic models can then be compared against each other by likelihood, with protection from the concern of inferring support for one model by comparison against a weak ``straw man'' alternative.
The use of benchmark likelihoods is demonstrated by \citep{king08,he10,wheeler23}.
Here, we use two benchmarks:

\noindent (i) A negative binomial model which is independent and identically distributed (IID) for each time point with a unit, with a unit-specific mean. We parameterize the model in terms of its mean and variance, as
\begin{equation}
\E[Y_{u,n}]=\mu_u\quad\text{and}\quad \var[Y_{u,n}] = \mu_u + \mu_u^2/s,
\end{equation}
where $s$ is a scale parameter.

\noindent (ii) An autoregressive negative binomial model, where the count $Y_{u,n}$ is modeled as negative binomial conditional on $Y_{u,n-1}$ with mean and variance given by
\begin{equation}
\E\big[Y_{u,n}|Y_{u,n-1}\big]=\mu_u + \phi Y_{u,n-1} \quad\text{and}\quad  \var\big[Y_{u,n}|Y_{u,n-1}\big] = \mu_u + \phi Y_{u,n-1} + \big(\mu_u+ \phi Y_{u,n-1}\big)^2/s,
\end{equation}
with the convention that $Y_{u,0}=0$.

We did not adopt the previously used log-ARMA benchmark, because it is inappropriate for count data with many zeros.
The likelihood was optimized using \code{optim} in R.

%%%%%%%%%%%% Review of inference methods %%%%%%%%%%%%%%

\section{\secSep Review of inference methods for metapopulation models}
\label{sec:review}

Viewing metapopulation models as high-dimensional structured population models, we consider the scalability of techniques developed for inferring population dynamics at a single location reviewed by \cite{funk20} and \cite{auger-methe21}.
These methods account for the nonlinear, stochastic, partially observed nature of biological dynamics available within the general class of POMP models.
Commonly implemented inference approaches for POMP models can be categorized as  (i) variants of MCMC; (ii) matching summary statistics between data and simulations; (iii)  linearization;   (iv) particle filters (i.e., sequential Monte Carlo).
We consider each of these in turn.

In principle, MCMC techniques enable Bayesian inference or maximum likelihood via expectation-maximization algorithms \cite{cappe05}.
In practice, successful MCMC for metapopulation models requires careful model-specific algorithm development \cite{whitehouse23}.

Matching summary statistics of simulations to the corresponding data statistic is, in principle, a readily applicable inference approach for a wide class of models including metapopulation models.
In the context of Bayesian inference this is called Approximate Bayesian Computing \cite{conlan12}.
However, informative, low-dimensional summary statistics can be hard to construct even for low-dimensional nonlinear systems.
This can make summary statistic methods statistically inefficient \cite{fasiolo16}.
This approach is practical when there is a small number of parameters to estimate and a large amount of data, in which case statistical efficiency may not be a concern.

Population dynamics may be approximately linear on a log scale, and this has been used to develop linearization methods for epidemiological time series analysis \cite{bjornstad02} that have been extended to metapopulation analysis \cite{xia04}.
This provides a numerically convenient set of tools, but requires scientists to work within a limited class of models.

Particle filter methods can provide statistically efficient inference for general POMP models but do not scale well with the dimension of the system.
This has led to the development of the block particle filter (BPF), described further in Section~\ref{sec:bpfilter}, and the ensemble Kalman filter (EnKF), discussed in Section~\ref{sec:enkf}.

%%%%%%%%%%%%%%%%%%%%%%%%% bpfilter and ibpf %%%%%%%%%%%%%%%%%%%%%%

\section{\secSep The block particle filter and iterated block particle filter}
\label{sec:bpfilter}

In Figure~{\mainFigIBPF} of the main text, we described BPF and the iterated block particle filter (IBPF). 
This section adds additional details on our implementation of these methods. 
We use the BPF of \citet{rebeschini15} implemented as \code{bpfilter} in the \pkg{spatPomp} package \citep{asfaw23arxiv}.
A filter can evaluate the likelihood function but is not directly concerned with parameter estimation.
IBPF algorithms extending BPF to enable parameter estimation were developed by \citet{ning23} and \citet{ionides22} and are implemented as \code{ibpf} in \pkg{spatPomp}.
Here, we give in informal introduction to \code{bpfilter} and \code{ibpf}.

The particle filter \citep{arulampalam02,doucet11} can be heuristically understood as Darwinian evolution operating on a swarm of particles.
Between consecutive observation times, each particle follows a random trajectory of the stochastic dynamic system.
This randomness is analogous to Darwinian mutation.
At an observation time, the particles are resamples with weights corresponding to the conditional density of the data given the location of the particle.
The weights are Darwinian fitness, and the resampling is Darwinian natural selection.

The performance of particle filters decays rapidly with the dimension of the latent state \citep{bengtsson08}.
Iterated particle filters suffer from the same curse of dimensionality.
Block particle filters avoid this curse by partitioning the latent states into weakly dependent units, and resampling separately on each unit.
This is analogous to recombination in sexual reproduction, with each block corresponding to a chromosome.
In this evolutionary analogy, each particle at time $t_n$ is an individual in the $n$th generation of a population.
The latent state of the particle is its genetic material, and this state is divided into chromosomes corresponding to each block.
Reproduction occurs at the resampling stage of the block particle filter, at which point the next generation of particles is resampled from the pool of chromosomes.
In this resampling process, each chromosome from the previous generation is selected proportionally to its fitness.
Thus, recombination allows successful blocks of one particle to join up with different successful blocks from another particle.
If this approximation permits effective high-dimensional filtering then the parameter perturbation strategy can be employed for parameter estimation, just as for basic particle filters.

\subsection{Log-likelihood estimation via filtering}
\label{sec:loglik:comparison}

For our primary goal of carrying out inference on unknown model parameters, the principal motivation for filtering is to obtain an estimate of the log-likelihood.
The log-likelihood is the probability density of the model, evaluated at the data, viewed as a function of the model parameters, defined as
\begin{equation}
\loglik(\theta)= \log\left(f_{Y_{1:N}}(y^*_{1:N} ; \theta)\right),
\end{equation}
This quantity is fundamental for likelihood-based inference, including maximum likelihood estimation and Bayesian inference (via combination of the likelihood with prior beliefs).

The recursive nature of a filtering algorithm suggests using a likelihood decomposition
\begin{equation}
f_{Y_{1:N}}(y^*_{1:N} ; \theta) = \prod_{n=1}^N f_{Y_n|Y_{1:n-1}}(y^*_n|y^*_{1:n-1} ; \theta).
\end{equation}
The requirement of a filtering algorithm is to provide an approximation to
\[
f_{X_n|Y_{1:n}}(x_n|y^*_{1:n} ; \theta),
\]
though many (including EnKF) also provide an approximation to the one-step prediction distribution, say
\[
f^P_{X_{n+1}|Y_{1:n}}(x_{n+1}|y^*_{1:n};\theta)\approx f_{X_{n+1}|Y_{1:n}}(x_{n+1}|y^*_{1:n} ; \theta).
\]
The  approximations $f^P_{X_{n+1}|Y_{1:n}}$ together with a measurement density can be used to construct a model defined by a joint probability density,
\begin{equation}\label{f^P}
f^P(y_{1:N};\theta)=\int \prod_{n=1}^N f_{Y_n|X_n}(y_n|x_n;\theta)
f^P_{n+1}(x_{n+1}|y_{1:n};\theta) \, dx_{1:N},
\end{equation}
with corresponding likelihood and log-likelihood functions,
\[
\lik^P(\theta) = f^P(y^*_{1:N};\theta), \hspace{1cm} \loglik^P(\theta)=\log \lik^P(\theta).
\]
Since $f^P_{X_{n+1}|Y_{1:n}}$ is an approximation, $\lik^P(\theta)$ does not exactly match the likelihood of the proposed model.
We call $\lik^P(\theta)$ the predictive likelihood of the filtering algorithm applied to the model.
It can be viewed as the exact likelihood of the approximate model for the data defined by, rather than the approximate likelihood of the original target model.
This perspective implies that, if the proposed model is correct, the expected value of $\loglik^P(\theta) $, viewed as a random function of $Y_{1:N}$, is lower than  $\loglik(\theta)$ since log-likelihood is a proper scoring rule \citep{gneiting07}.
Therefore, it is appropriate to compare filtering methods by their predictive likelihoods.

\begin{figure}
\begin{center}
  \includegraphics[width=15cm]{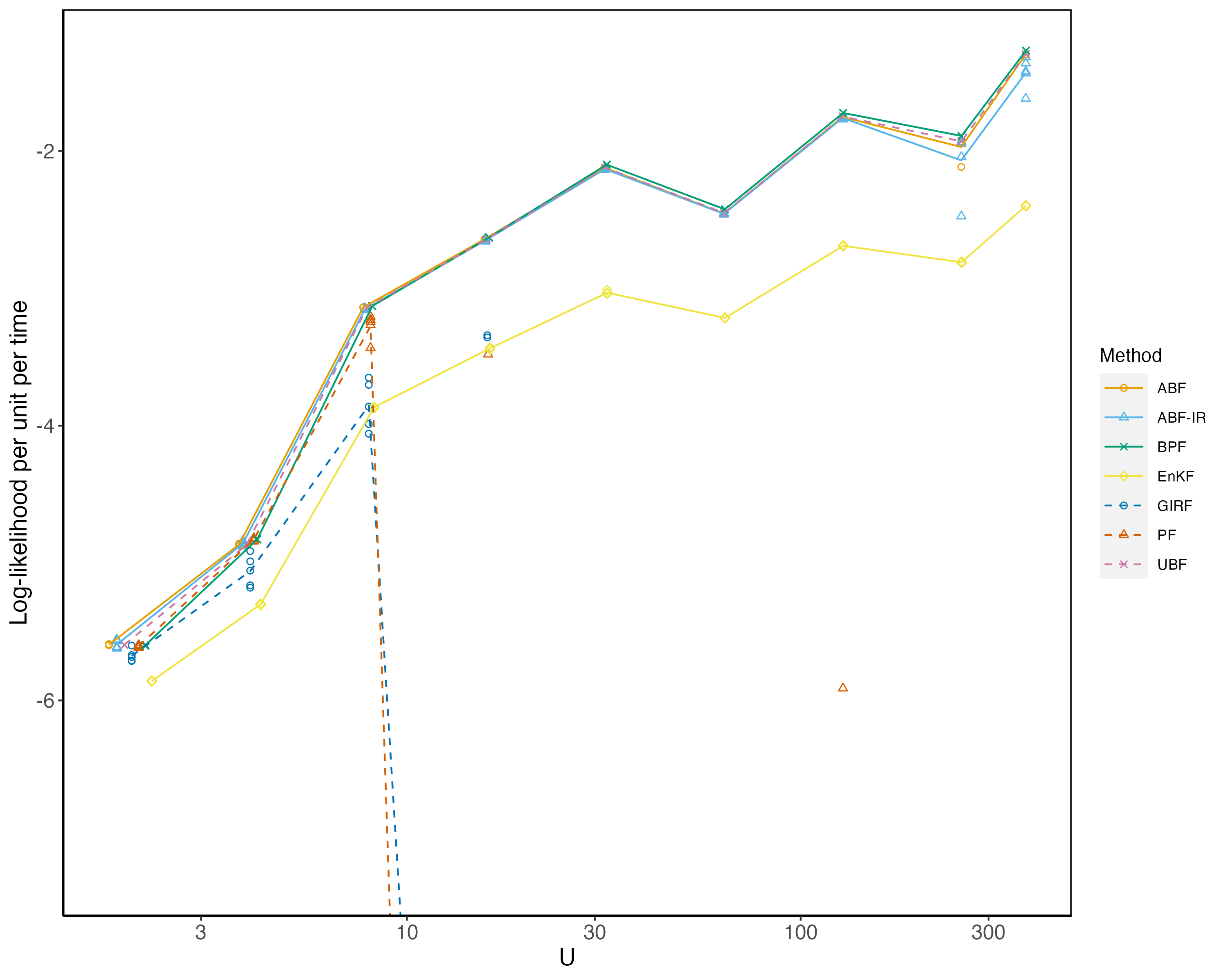}
\end{center}
\caption{Comparing filters on simulated data from model \code{li23}. ABF is the adapted bagged filter \citep{ionides23-jasa}; ABF-IR is the adapted bagged filter with intermediate resampling \citep{ionides23-jasa}; BPF is the block particle filter; EnKF is the ensemble Kalman filter; GIRF is the guided intermediate resampling filter \citep{park20}; PF is the particle filter; UBF is the unadapted bagged filter \citep{ionides23-jasa}.}\label{fig:filter_test}
\end{figure}

To investigate the suitability of the block particle filter for this data analysis, we compared log-likelihood evaluation from various filters available in the \pkg{spatPomp} package.
We used simulated data from our maximum likelihood estimate so that the comparison is made on a correctly specified model, because we do not want to assess filters based on how well they compensate for model misspecification.
For a correctly specified model, the best possible expected log-likelihood arises for an ideal filter; in other words, log-likelihood is a proper scoring method for filters \citep{gneiting07}.
Figure~\ref{fig:filter_test} shows that the block particle filter performs well on this task.
The particle filter is effective for up to $U=5$ cities, but (as theory predicts) its performance starts declining rapidly as dimension increases.
The particle filter gives an unbiased and consistent Monte Carlo estimate of the likelihood, so we can tell from Figure~\ref{fig:filter_test} that the block approximation is effective for small $U$ since is it does not fall far behind the particle filter in situations where the latter is known to give essentially exact results.
However, the block particle filter continues to operate successfully for large $U$, when the particle filter fails.
Two bagged filters (ABF and UBF) also perform well, however their structure is not well suited to parameter estimation via iterated perturbed filtering \citep{ionides23-jasa}.
The particle filters using guided intermediate resampling (ABF-IR and GIRF) are computationally expensive; in principle, they have good scalability properties, however, their \pkg{spatPomp} implementation performs less well than the unguided filters in this experiment.

The algorithmic settings for Figure~\ref{fig:filter_test} were set for comparable computing times in the \pkg{spatPomp} implementation.
Full details are available in the published source code.
Different algorithms have different demands in terms of number of computations, memory requirements, and parallelizability.
All these algorithms have various tuning parameters, so we do not rule out the possibility that the numerical comparisons are dependent on details of the implementation.
Figure~\ref{fig:filter_test} is similar to Figure~3 of \citet{ionides23-jasa}, that shows an equivalent filter comparison for a longer collection of time series of an endemic viral disease, namely pre-vaccination measles.
Different relative performance results can be obtained on other model classes, for example in a linear Gaussian model (Figure~1 of \citet{ionides23-jasa}) or the Lorenz~96 geophysical model (Figure~S3 of \citet{ionides23-jasa}).
Figure~\ref{fig:filter_test} therefore adds to the growing body of evidence that block particle filters are well suited to metapopulation models.

\subsection{\secSep Limitations of IBPF}

Simulation-based methods are computationally intensive.
IBPF requires thousands of simulations in each of hundreds of filtering iterations.
In practice, this limits the applicability to a moderate scale, say, hundreds of units.

IBPF approximates a full-information maximum-likelihood analysis, yet both the likelihood evaluation and maximization are subject to errors that could become scientifically significant.
IBPF builds on the block particle filter, and so IBPF cannot expect to succeed in situations where BPF fails.
However, it is possible for IBPF to fail in situations where BPF succeeds, 
We discuss separately these two requirements for success of IBPF.

BPF has good scalability properties, but an approximation error which can be large when the model is not a good match for the method.
The theory within which BPF has small approximation error assumes that dependence decays suitably quickly with distance between units \citep{rebeschini15}.
The theory for IBPF, in the case where all parameters are unit-specific, has a similar requirement, but on an extended model where the latent state at each unit is augmented with a parameter vector carrying out a random walk \citep{ning23}.
There is not yet a comparable theory for the shared parameter extension of IBPF proposed and demonstrated by \citet{ionides22}, but a similar requirement is probably needed.

The size of an initial infected population at a single source unit, Wuhan for COVID-19, is an example of a potentially problematic parameter to estimate via IBPF.
In the extended model, this parameter is local to the state at Wuhan, and yet it is important for the dynamics of other units.
Fortunately, the initialization procedure does award this parameter some direct effect on the initial number of infections in other units, which the algorithm can harness.
Without this, the IBPF algorithm would lose any power for events outside the block containing Wuhan to inform the initial state in Wuhan.

Determining the success of a filter, in a particular application, is an empirical task.
Many filters, including BPF and the basic particle filter, evaluate the log-likelihood by constructing a sequence of one-step predictive distributions which do not have access to future data.
The log-likelihood is a proper scoring rule for such forecasts \citep{gneiting07}, meaning that, if the model is correct, no other predictive distribution can have higher expected log-likelihood.
This motivates comparison of filters by their estimated log-likelihood: the higher, the better.

Caution is required in the presence of model misspecification.
For example, the model may place zero probability on specific latent state values, yet these states may become possible in a block particle filter due to the block resampling.
If the data favor these inconsistent values (an indication of model misspecification) then the block particle filter may have much higher likelihood that an ideal filter.
For this reason, we recommend that experimentation to determine the choice of filter should be carried out on simulated data as well as the actual data.
If the conclusions are different in these two scenarios, that is an indication of model misspecification.

%%%%%%%%%%%%%%%%%%%%% enkf and ienkf %%%%%%%%%%%%%%%%%%

\section{\secSep The ensemble Kalman filter (EnKF) and its use for metapopulation models}
\label{sec:enkf}

\begin{algorithm}[tb]
  \caption{EnKF algorithm \citep[adapted from][]{asfaw23arxiv}}\label{alg:enkf}
  \KwIn{simulator for the transition density, $f_{\myvec{X}_{\ttime}|\myvec{X}_{\ttime-1}}(\myvec{x}_{\ttime} \given \myvec{x}_{\ttime-1}\giventh\theta)$, and initial density, $f_{\myvec{X}_0}(\myvec{x}_0\giventh\theta)$;
    evaluator for expectation of $Y_{\unit,\ttime}$ given $X_{\unit,\ttime}=x$, ${\emeasure}_{\unit}(x,\theta)$, and corresponding variance, ${\vmeasure}_{\unit}(x,\theta)$;
    parameter, $\theta$;
    data, $\data{\myvec{y}}_{1:\Time}$;
    number of particles, $J$.
  }
  initialize filter particles,
  $\myvec{X}_{0}^{F,\np}\sim {f}_{\myvec{X}_{0}}\left(\mydot\giventh{\theta}\right)$
  for $\np$ in $\seq{1}{\Np}$
  \;
      \For{$\ttime\ \mathrm{in} \ \seq{1}{\Time}$}{
        prediction ensemble,
    $\myvec{X}_{\ttime}^{P,\np}\sim {f}_{\myvec{X}_{\ttime}|\myvec{X}_{\ttime-1}}\big(\mydot|\myvec{X}_{\ttime-1}^{F,\np};\theta\big)$
    for $\np$ in $\seq{1}{\Np}$
    \nllabel{alg:enkf:prediction}
    \;
        centered prediction ensemble, $\tilde{\myvec{X}}_{\ttime}^{P,\np} =
        \myvec{X}_{\ttime}^{P,\np} - \frac{1}{\Np}\sum_{\altNp=1}^{\Np}\myvec{X}_{\ttime}^{P,\altNp}$
       for $\np$ in $\seq{1}{\Np}$
    \;
        forecast ensemble, $\myvec{\hat{Y}}^{\np}_{\!\ttime}={\emeasure}_{\unit}(X_{\unit,\ttime}^{P,\np},\theta)$
       for $\np$ in $\seq{1}{\Np}$
        \nllabel{alg:enkf:forecast}
        \;
	forecast mean, $\overline{\myvec{Y}}_{\!\ttime}=\frac{1}{\Np}\sum_{\np=1}^{\Np}\myvec{\hat{Y}}^{\np}_{\!\ttime}$
	\;
        centered forecast ensemble, $\myvec{\tilde{Y}}^{\np}_{\ttime} =
        \myvec{\hat{Y}}^{\np}_{\!\ttime} - \overline{\myvec{Y}}_{\!\ttime}$
       for $\np$ in $\seq{1}{\Np}$
        \;
        forecast measurement variance,
	$R_{\unit,\altUnit} = \mathbbm{1}_{\unit,\altUnit} \,
	  \frac{1}{\Np}\sum_{\np=1}^{\Np}
	    {\vmeasure}_{\unit}\big(
	      \myvec{X}_{\unit,\ttime}^{P,\np},\theta\big)$
         \nllabel{alg:enkf:cond:var}
	 for $\unit, \altUnit$ in $\seq{1}{\Unit}$
        \;
        forecast estimated covariance, $\Sigma_{Y}= \frac{1}{\Np-1}\sum_{\np=1}^{\Np}(\myvec{\tilde{Y}}^{\np}_{\!\ttime})(\myvec{\tilde{Y}}^{\np}_{\!\ttime})^T + R$
        \;
        prediction and forecast sample covariance, $\Sigma_{XY}=\frac{1}{\Np-1}\sum_{\np=1}^{\Np}(\tilde{\myvec{X}}_{\ttime}^{P,\np})(\myvec{\tilde{Y}}^{\np}_{\!\ttime})^T$
        \;
        Kalman gain, $K = \Sigma_{XY}\Sigma_{Y}^{-1}$
	\nllabel{alg:enkf:kalman_gain}
        \;
        artificial measurement noise, $\myvec{\epsilon}_{\ttime}^{\np}\sim \normal(\myvec{0},R)$
	for $\np$ in $\seq{1}{\Np}$
        \nllabel{alg:enkf:artificial:noise}
        \;
        errors, $\myvec{r}_{\ttime}^{\np}= \myvec{\hat{Y}}^{\np}_{\!\ttime} - \data{\myvec{y}}_{\ttime}$
	for $\np$ in $\seq{1}{\Np}$
        \;
        filter update,
        $\myvec{X}_{\ttime}^{F,\np} = \myvec{X}_{\ttime}^{P,\np} +
              K\big( \myvec{r}_{\ttime}^{\np}+\myvec{\epsilon}_{\ttime}^{\np}\big)$
	for $\np$ in $\seq{1}{\Np}$
        \nllabel{alg:enkf:update}
	\;
	$\loglik_{\ttime}=\log \big[ \phi\big(\data{\myvec{y}}_{\ttime} \giventh \overline{\myvec{Y}}_{\!\ttime} , \Sigma_{Y} \big) \big]$ where $\phi(\cdot\giventh \myvec{\mu},\Sigma)$ is the $\normal(\myvec{\mu},\Sigma)$ density.
      }
  \KwOut{
    filter sample, $\myvec{X}^{F,1:\Np}_{\ttime}$, for $n$ in $\seq{1}{N}$;
    log likelihood estimate, $\loglik^{\mbox{\tiny{EnKF}}}=\sum_{\ttime=1}^{\Time} \loglik_{\ttime}$
  }
\end{algorithm}

EnKF algorithms \citep{evensen22} have proved effective for moderately nonlinear, non-Gaussian data assimilation tasks with large amounts of data.
Algorithm~\ref{alg:enkf} gives pseudocode for an EnKF algorithm, using notation for SpatPOMP models consistent with the \pkg{spatPomp} R package \citep{asfaw23arxiv}.
EnKF algorithms update each member of an ensemble using a Kalman gain, constructed in line~\ref{alg:enkf:kalman_gain} of Algorithm~\ref{alg:enkf}.
This linear update rule corresponds to an ideal filter (i.e., the Kalman filter) when the observations and latent states are jointly linear and Gaussian.
The nonlinear and non-Gaussian behavior permitted in the prediction step (line~\ref{alg:enkf:prediction}) makes some appropriate adaptation for general SpatPOMP models, but does not in general guarantee a good approximation to the ideal nonlinear filter.
We see in Figure~\ref{fig:filter_test} that the performance of EnKF on the \code{li23} model falls substantially below some filters with nonlinear update rules.

A technical requirement for proper likelihood-based comparison of two models is that the likelihoods are calculated with respect to the same base measure.
This technical consideration can become important in the context of EnKF since this algorithm is motivated by a continuous, real-valued probability distribution (the Gaussian distribution) yet is applied to discrete, integer-valued metapopulation models.
When population counts are not small, evaluating a Gaussian probability density function at integer values may be a close approximation to the probability mass function of a discrete model.
However, when counts are small, we may encounter a situation where the prediction variance is small, in which case the Gaussian probability density is unbounded.
By contrast, a valid probability mass function (i.e., a discrete probability density with respect to counting measure) can never attain values higher than 1.

\citet{li20} prevented this issue by putting a lower bound of $4$ on the observation variance for their EnKF implementation.
In addition, they added addition variance to their EnKF estimate, discussed further in Section~\ref{subsec:enkf-mismatch}. 
We implemented their approach in our EnKF calculations for Figure~\ref{fig:filter_test} by replacing $V_{\unit,n}$ in equation (\ref{eq:vunit_measure}) with
\begin{equation}
\label{eq:vunit_measure_enkf}
V_{\unit,n}=\min(4,C_{\unit,n}^2/4)
\end{equation}
when carrying out inference via the ensemble Kalman filter.
This is similar to taking $\tau=0.5$ in (\ref{eq:vunit_measure}).
By contrast, we set $\tau=0$ for model {\LiMobility} in Table~\ref{tab:parameters}, since that corresponds to the parameter value used by \citet{li20} to study the properties of the fitted model.

Further, \citet{li20} introduce a measure of discrepancy between the fitted model and the data which they call log-likelihood but which is not an approximation to the statistical quantity $\lik(\theta)$.
The quantity shown in their Figure~1, and described in their supplementary Section~8, could be viewed as an approximation to a pseudo-log-likelihood \citep{besag74}.
However, this quantity is not the predictive log-likelihood of any model, and cannot properly be compared with log-likelihoods from alternative models.
An alternative approach for employing EnKF algorithms with discrete data is to embed EnKF within a Markov chain Monte Carlo algorithm \citep{katzfuss19}.

\subsection{Mismatches between the model and the EnKF specification}
\label{subsec:enkf-mismatch}

Modifications to the EnKF implementation may improve filtering but break the correspondence between the algorithm and the postulated model.
For example, \citep{li20} use a measurement variance that depends on the observation itself, which superficially suggests the mathematically inconsistent expression
\[
\var(Y_n|X_n) = \min(4,Y_n^2/4).
\]
Filtering may proceed with this variance specification, but it does not correspond to a valid predictive distribution since a conditional variance for $Y_n$ cannot depend on $Y_n$.

Some mismatch between the predictive likelihood and the actual model likelihood occurs whenever using numerical methods.
Inference is necessarily based on the numerically implementation of the filtered model and its likelihood, implying that the predictive likelihood is a proper measure of fit for the procedure actually implemented.
As pointed out in Sec.~\ref{sec:loglik:comparison}, a filter generating a legitimate predictive likelihood is penalized for infelicity to the intended model, so far as the intended model fits the data.
This permits likelihood-based comparison candidate filters as well as candidate models.
However, a non-predictive likelihood approximation calculated requires additional care in its interpretation; use of future information could lead to higher likelihoods than can be obtained by even an ideal filter.
If it is expedient to use a non-predictive likelihood approximation, this issue requires care. 

%\clearpage

\section{\secSep Estimation for the unconstrained model, {\RevisedModelUnconstrained}}
\label{sec:unconstrained_estimation}

\begin{knitrout}
\definecolor{shadecolor}{rgb}{0.969, 0.969, 0.969}\color{fgcolor}\begin{figure}

{\centering \includegraphics[width=5in]{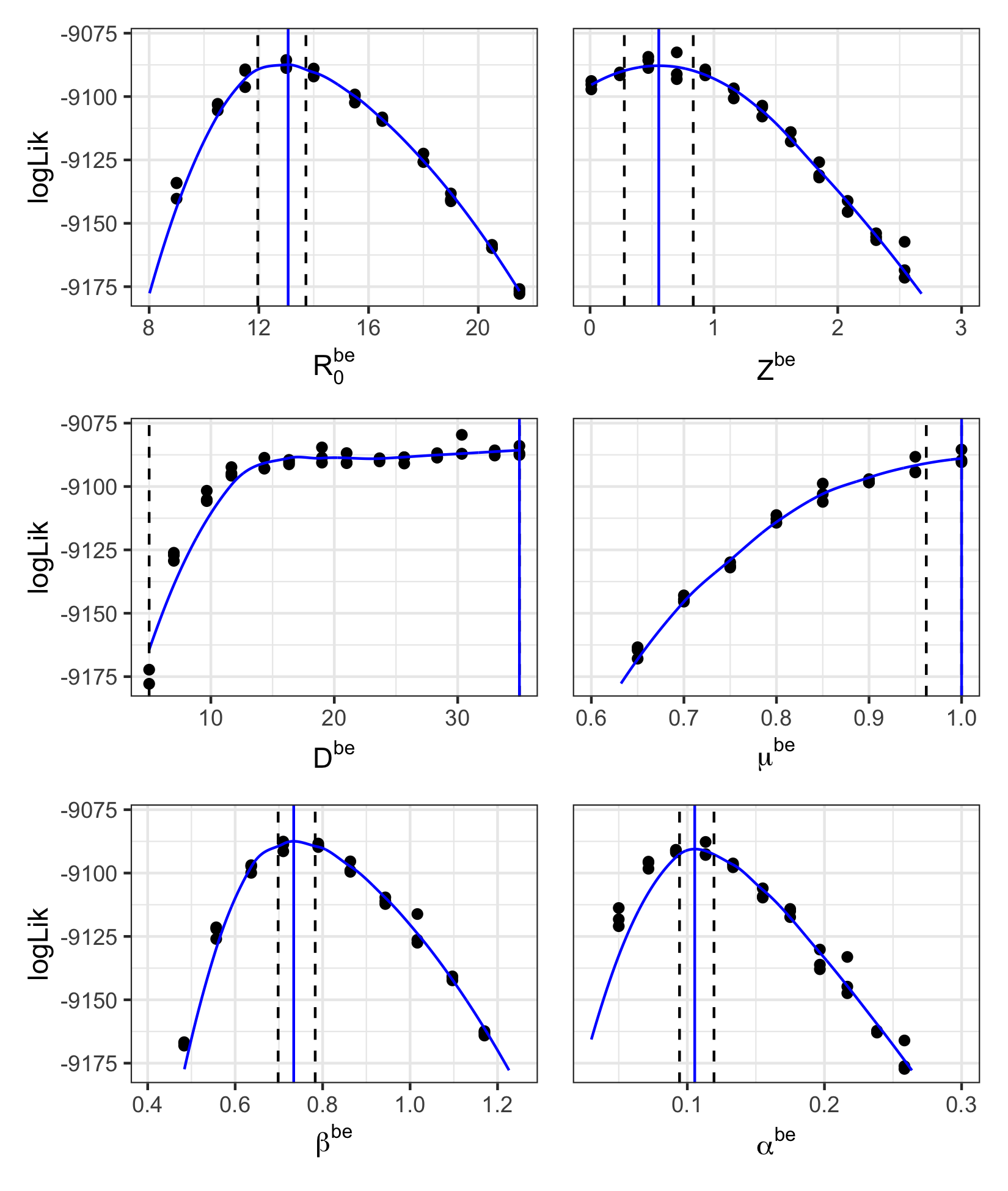} 

}

\caption{Profile log-likelihood for model {\RevisedModelUnconstrained} parameters before lockdown: $\mathcal{R}_0^{\before}$, $Z^{\before}$, $D^{\before}$, $\mu^{\before}$ and $\beta^{\before}$.}\label{fig:be_profiles_uc}
\end{figure}

\end{knitrout}

\begin{knitrout}
\definecolor{shadecolor}{rgb}{0.969, 0.969, 0.969}\color{fgcolor}\begin{figure}

{\centering \includegraphics[width=5in]{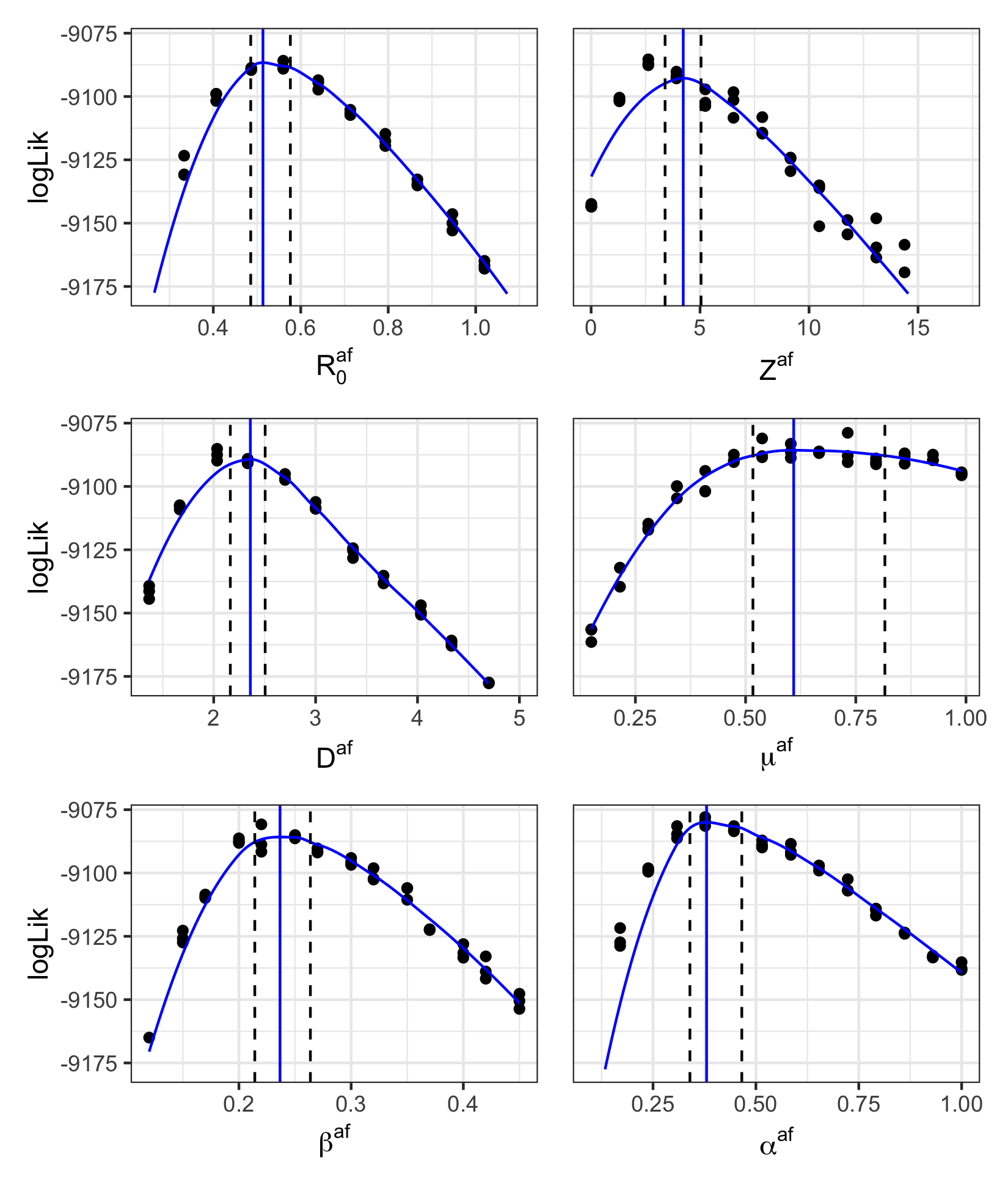} 

}

\caption{Profile log-likelihood for model {\RevisedModelUnconstrained} parameters after lockdown: $\mathcal{R}_0^{\after}$, $Z^{\after}$, $D^{\after}$, $\mu^{\after}$, $\beta^{\after}$ and $\alpha^{\after}$.}\label{fig:af_profiles_uc}
\end{figure}

\end{knitrout}

\begin{knitrout}
\definecolor{shadecolor}{rgb}{0.969, 0.969, 0.969}\color{fgcolor}\begin{figure}

{\centering \includegraphics[width=5in]{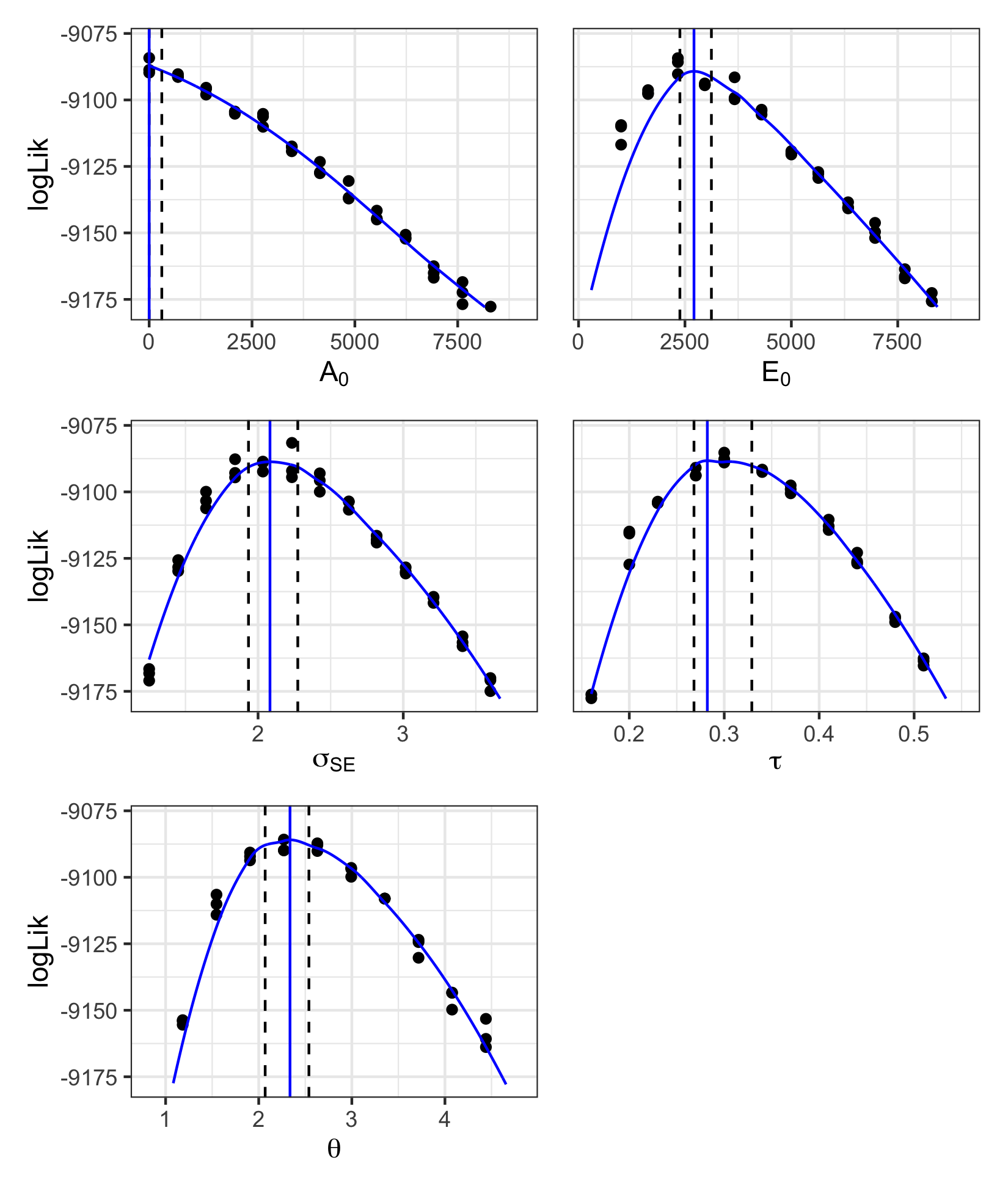} 

}

\caption{Profile log-likelihood for model {\RevisedModelUnconstrained} parameters which are unaffected by lockdown: $\tau$, $\theta$, $\sigma_{SE}$, $A_0$ and $E_0$.}\label{fig:constant_profiles_uc}
\end{figure}

\end{knitrout}

We calculated profile likelihoods for each parameter in {\RevisedModelUnconstrained}, in order to assess identifability and obtain confidence intervals.
Profile likelihood involves fixing one parameter at a range of values while maximizing the log-likelihood with respect to all the other estimated parameters.
We use Monte Carlo adjusted profile (MCAP) methodology which provides a way to construct likelihood-based confidence intervals in situations where Monte~Carlo variability in maximization and evaluation of the log-likelihood is too large to ignore \citep{ionides17,ning21}.
An estimate of the profile likelihood is obtained by applying a smoothing algorithm (such as a smoothing spline) to these noisy evaluations.
The MCAP confidence interval selects the region of the estimated profile above a cutoff value, where the cutoff is chosen to combine the statistical uncertainty of the ideal (inaccessible) likelihood function with the Monte Carlo variability of the available estimate of the likelihood function.
MCAP is not appropriate when the maximum likelihood occurs on the boundary of the parameter space, which occurs here for the initial unobserved cases, $A_0$, and the initial relative transmissibility, $\mu^{\before}$.
For these parameters we therefore used a basic likelihood ratio test on the smoothed likelihood, unadjusted for Monte Carlo error.

The resulting profiles are shown in Figures~\ref{fig:be_profiles_uc},~\ref{fig:af_profiles_uc} and~\ref{fig:constant_profiles_uc}.
We see from Figure~\ref{fig:be_profiles_uc} that $\infectiousPeriod^{\before}$ is weakly identified, and arbitrarily high values of this parameter can be consistent with the data.
In this model, the mean 9-day distributed delay in reporting  means that the initial dynamics cannot have many visible consequences in the early data.
This is consistent with the absence of reported cases in the first 6 days of the dataset.
However, as a consequence, the data have limited ability to identify model parameters, increasing the possibility that certain parameters, or combinations of parameters, have large statistical uncertainty.
We resolved this situation by adding two constraints, $\infectiousPeriod^{\before}=\infectiousPeriod^{\after}=\infectiousPeriod$ and $\latency^{\before}=\latency^{\after}=\latency$, to obtain the constrained model, {\RevisedModelConstrained}.

%\clearpage

\section{\secSep Estimation for the constrained model, {\RevisedModelConstrained}}
\label{sec:constrained_estimation}

\begin{knitrout}
\definecolor{shadecolor}{rgb}{0.969, 0.969, 0.969}\color{fgcolor}\begin{figure}

{\centering \includegraphics[width=5in]{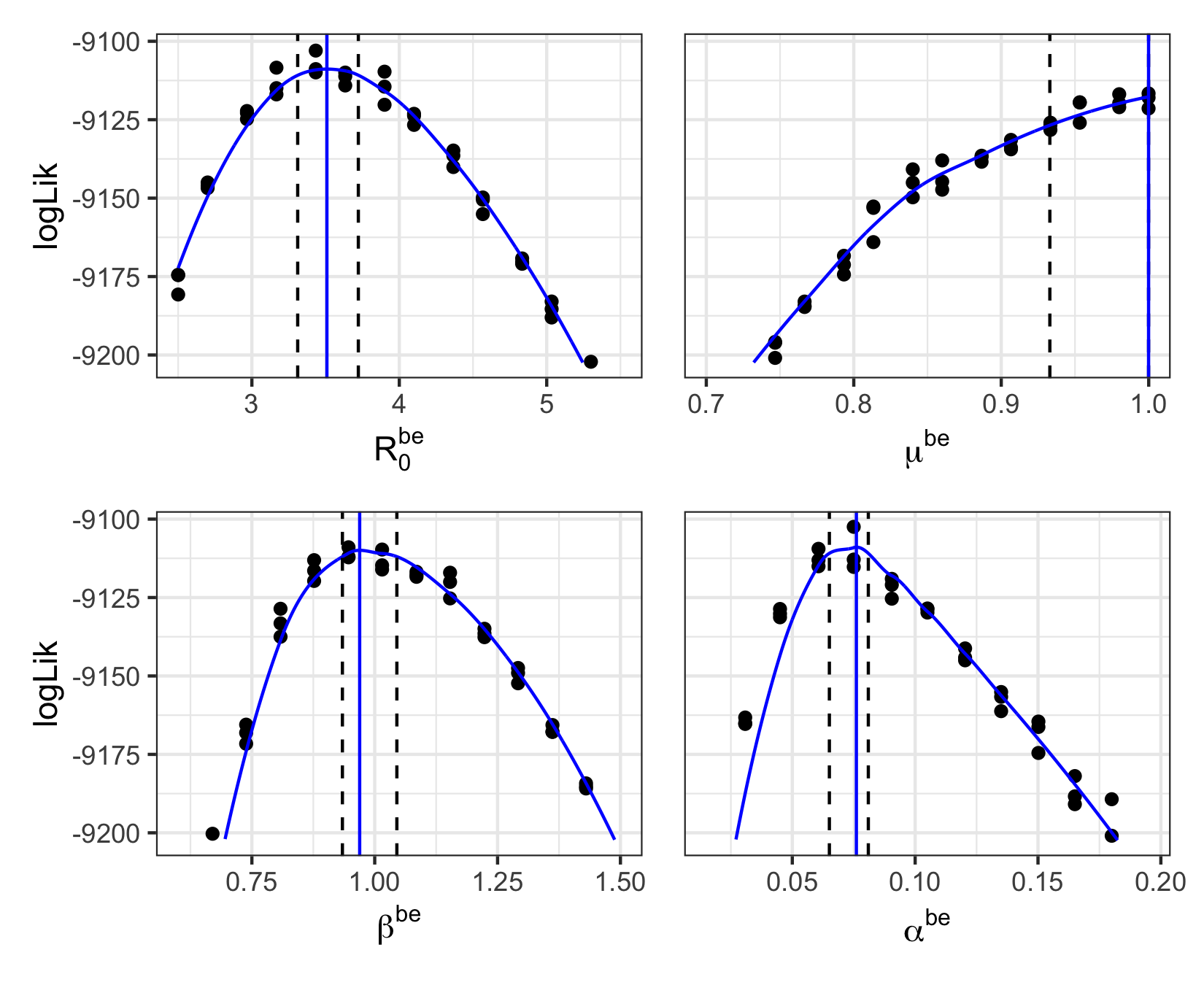} 

}

\caption{Profile log-likelihood for model {\RevisedModelConstrained} parameters before lockdown: $\mathcal{R}_0^{\before}$, $\mu^{\before}$ and $\beta^{\before}$.}\label{fig:be_profiles_c}
\end{figure}

\end{knitrout}

\begin{knitrout}
\definecolor{shadecolor}{rgb}{0.969, 0.969, 0.969}\color{fgcolor}\begin{figure}

{\centering \includegraphics[width=5in]{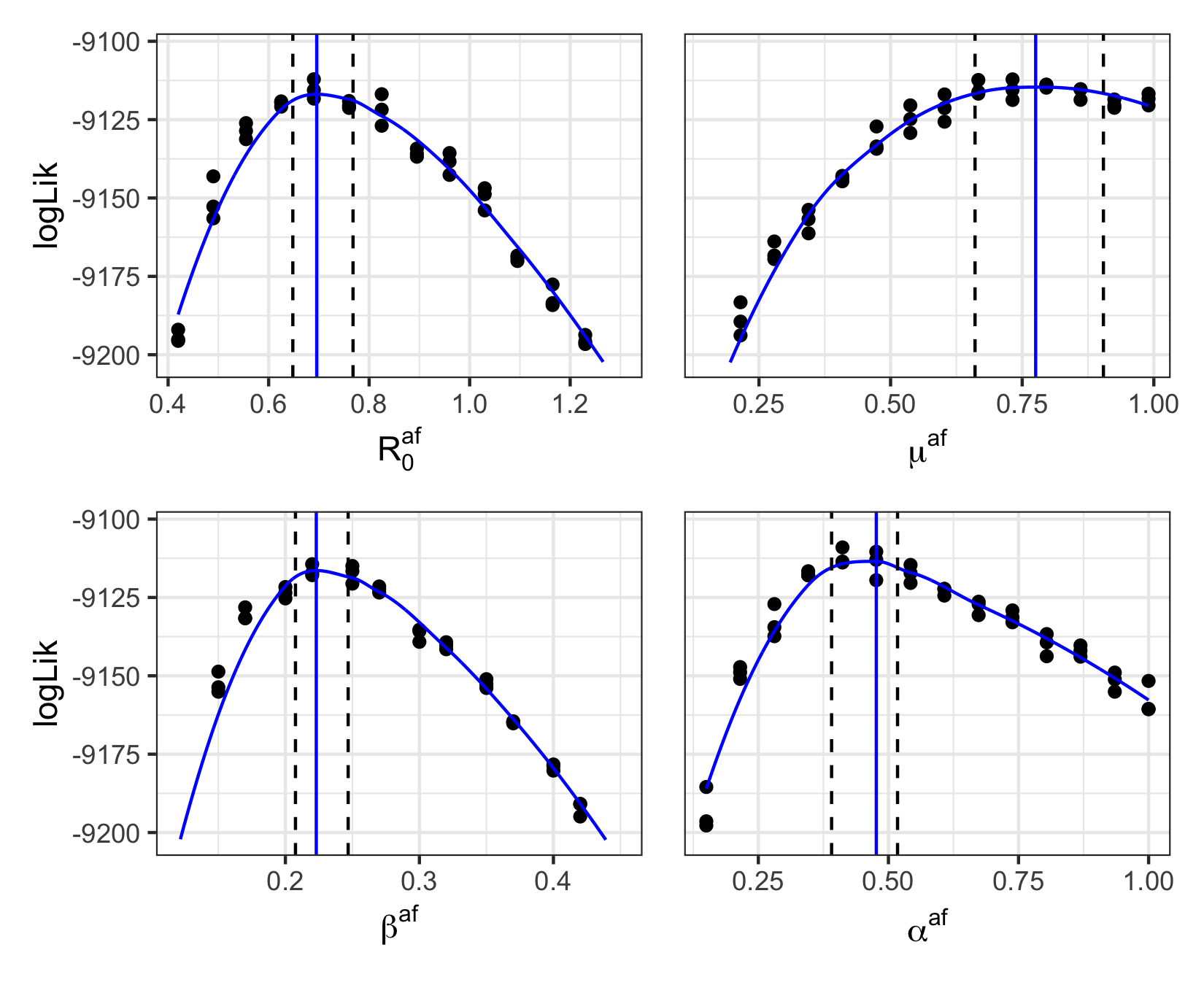} 

}

\caption{Profile log-likelihood for model {\RevisedModelConstrained} parameters after lockdown: $\mathcal{R}_0^{\after}$, $\mu^{\after}$, $\beta^{\after}$ and $\alpha^{\after}$ in model {\RevisedModelConstrained}}\label{fig:af_profiles_c}
\end{figure}

\end{knitrout}

\begin{knitrout}
\definecolor{shadecolor}{rgb}{0.969, 0.969, 0.969}\color{fgcolor}\begin{figure}

{\centering \includegraphics[width=5in]{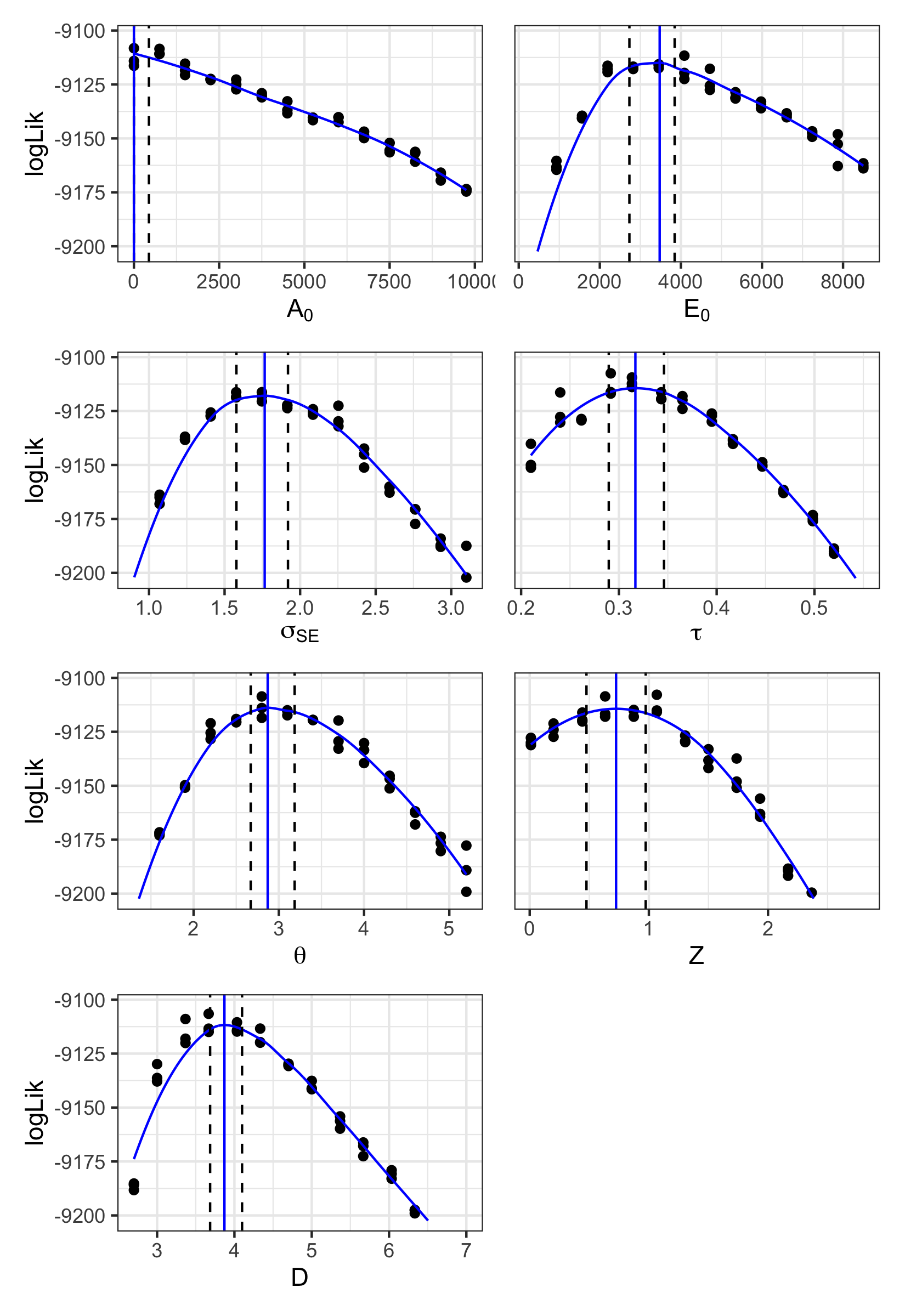} 

}

\caption{Profile log-likelihood for  model {\RevisedModelConstrained}  parameters which are unaffected by lockdown: $\tau$, $\theta$,  $Z$, $D$, $\sigma_{SE}$,  $A_0$ and $E_0$.}\label{fig:constant_profiles_c}
\end{figure}

\end{knitrout}

Figures~\ref{fig:be_profiles_c},~\ref{fig:af_profiles_c} and~\ref{fig:constant_profiles_c} graph the profile likelihood evaluations and construct the resulting confidence intervals, following the same procedures used for Figures~\ref{fig:be_profiles_uc},~\ref{fig:af_profiles_uc} and~\ref{fig:constant_profiles_uc}.
For both models~{\RevisedModelUnconstrained} and~{\RevisedModelConstrained}, the initial count of unreportable infections in Wuhan, $A_0$, is indistinguishable from $A_0=0$.
Evidently, the model prefers to explain the data by placing the initial cases in the latent state, $E$.
However, the evidence is not strong: the profiles for $A_0$ show compatibility with $A_0=5000$ for a cost of around 25-50 units of log likelihood.
That is strong statistical evidence in the context of the model under investigation, since a 95\% confidence interval contains only values within around 2 log units.

Formally, confidence intervals (like other forms of model-based statistical inference) are constructed based on a class of models under consideration.
The meaning of these intervals in the context of models outside the class under consideration is generally unclear.
However, the flatter the profile, the easier for some relatively small, unmodeled phenomenon to affect the estimate.
Thus, to understand the robustness of the results to model misspecification, it can be helpful to consider the effect of larger likelihood cutoffs.

Standard robust statistical methods concern proper inference when aspects of the model are statistically inadequate, but comparison against appropriate benchmarks provides protection against this type of model misspecification.
For example, we do not have to be excessively concerned about the possible effect of inappropriate modeling of dependence on confidence intervals if our mechanistic model has a likelihood comparing favorably against associative models having flexible specification of dependence. 
A different type of model misspecification arises when important explanatory variables are missing, or the postulated causal structures in the model class do not adequately represent reality.
Such unknowns cannot readily be accounted for in standard error estimates.
The curvature of the profile likelihood may indicate how robust the results are to small misspecifications.
For large misspecifications, parameters in differing models may have entirely different causal interpretation, and the respective likelihoods provide a measure of support from the data concerning each hypothesis.

%\clearpage

\section{\secSep Anomaly analysis}

\begin{knitrout}
\definecolor{shadecolor}{rgb}{0.969, 0.969, 0.969}\color{fgcolor}\begin{figure}

{\centering \includegraphics[width=6.5in]{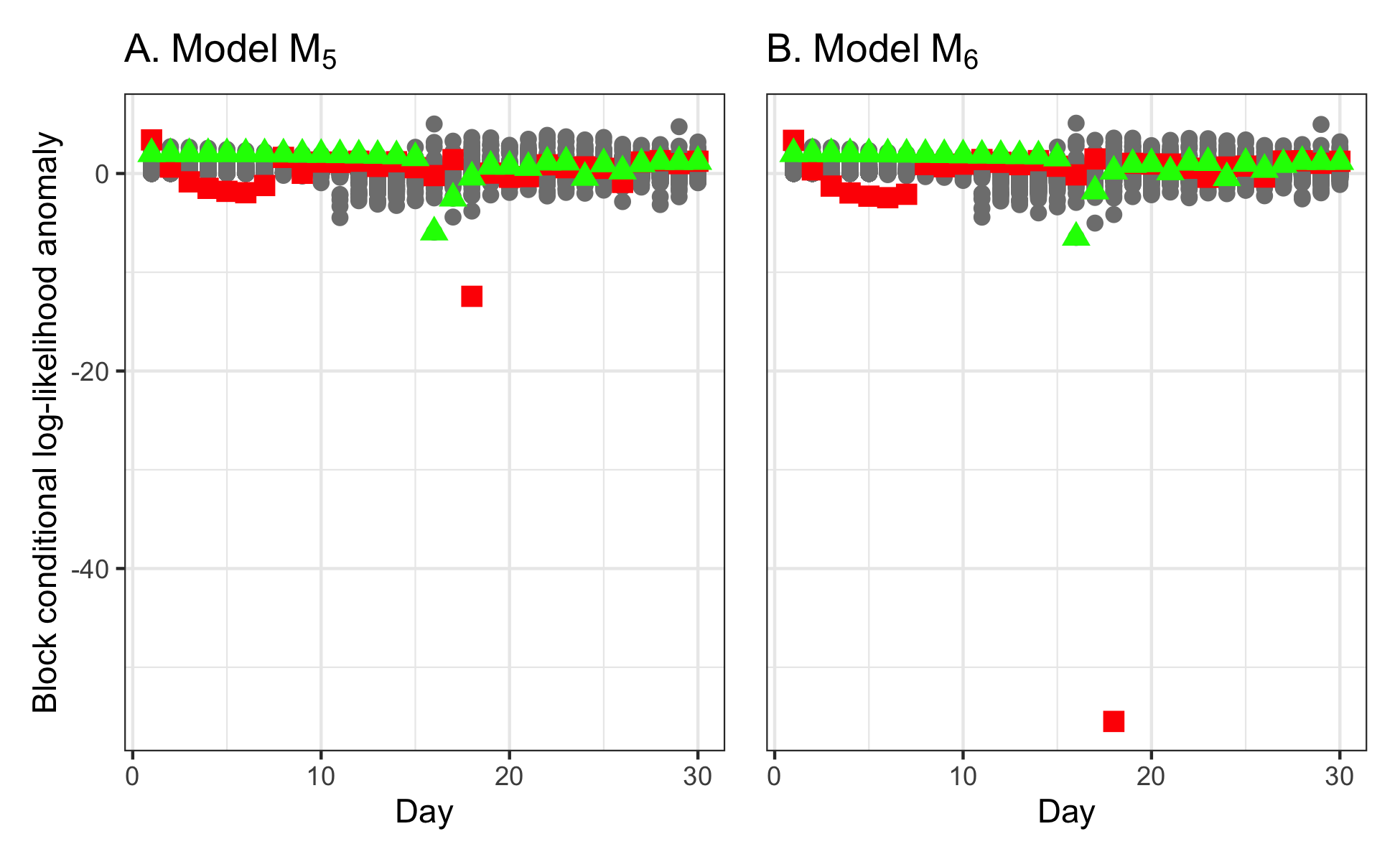} 

}

\caption[Conditional log-likelihood anomaly for each city at each time point]{Conditional log-likelihood anomaly for each city at each time point. Panel A corresponds to the best fit for {\RevisedModelUnconstrained} and panel B the best fit for {\RevisedModelConstrained}. Points for Wuhan are red squares, and Huangshi are green triangles.}\label{fig:residual-plot}
\end{figure}

\end{knitrout}

The block particle filter log-likelihood estimate can be decomposed into block conditional log-likelihoods for each city at each time point.
The total estimated log-likelihood for the full dataset is the sum of these block conditional log-likelihoods.
Likelihood depends on the units of the measured quantity, leading to a scale-dependent additive constant in the log-likelihood.
To remove this constant, and to compare the model with a simple statistical prediction, we consider the log-likelihood anomaly, defined to be the model log-likelihood minus the benchmark log-likelihood.
The log-likelihood anomaly at each time for each block is the corresponding difference for the block conditional log-likelihood.
These log-likelihood anomalies can be investigated to look for patterns of interest; they are analogous to the residuals (i.e., observations minus predicted values) used for diagnostic analysis of regression models.
An anomaly for an observation much smaller than $-1$ suggests that the data point is poorly explained by the mechanistic model, in which case it is called an outlier.

The largest outlier for both models~{\RevisedModelUnconstrained} and~{\RevisedModelConstrained} arises for Wuhan on day 18.
This corresponds to the dramatic increase in reported cases on that day, shown in Figure~\ref{fig:anomalous-cities-plot}.
This feature is a much more sever anomaly for the constrained model, {\RevisedModelConstrained}.
Indeed, the difference in the anomalies, which is $43.0$ log-likelihood units, is more than enough to explain the difference in maximized log-likelihood between these two models.
Evidently, the dynamics of the fitted model {\RevisedModelUnconstrained} manage to better explain this outlier by postulating a long latency period and high $\Rzero$ before the lockdown so that there is a larger supply of cases ready to explain the dramatic increase in reported cases on day 18.
We see that one extreme outlier, which may represent an idiosyncrasy of the reporting process rather than a feature of the underlying dynamics, can have substantial consequences on the fit and the resulting conclusions.
Essentially, the reported Wuhan cases on day 18 are inconsistent with model~{\RevisedModelConstrained}; it pays a heavy price for this in terms of log-likelihood, but that may not be a scientific concern since it may indeed be the case that the reported peak on day 18 was not a genuine feature of the dynamics.
Diagnostic protocols for COVID-19 were in their infancy, and changing rapidly, so the anomaly can be explained as a unique event in the reporting system.

The second largest outlier occurred on day 16 in Huangshi, a city only 100km from Wuhan.
Again, this corresponds to a sudden surge in reported cases (shown in Figure~\ref{fig:anomalous-cities-plot}) beyond what the model can account for. 

\begin{knitrout}
\definecolor{shadecolor}{rgb}{0.969, 0.969, 0.969}\color{fgcolor}\begin{figure}

{\centering \includegraphics[width=6.5in]{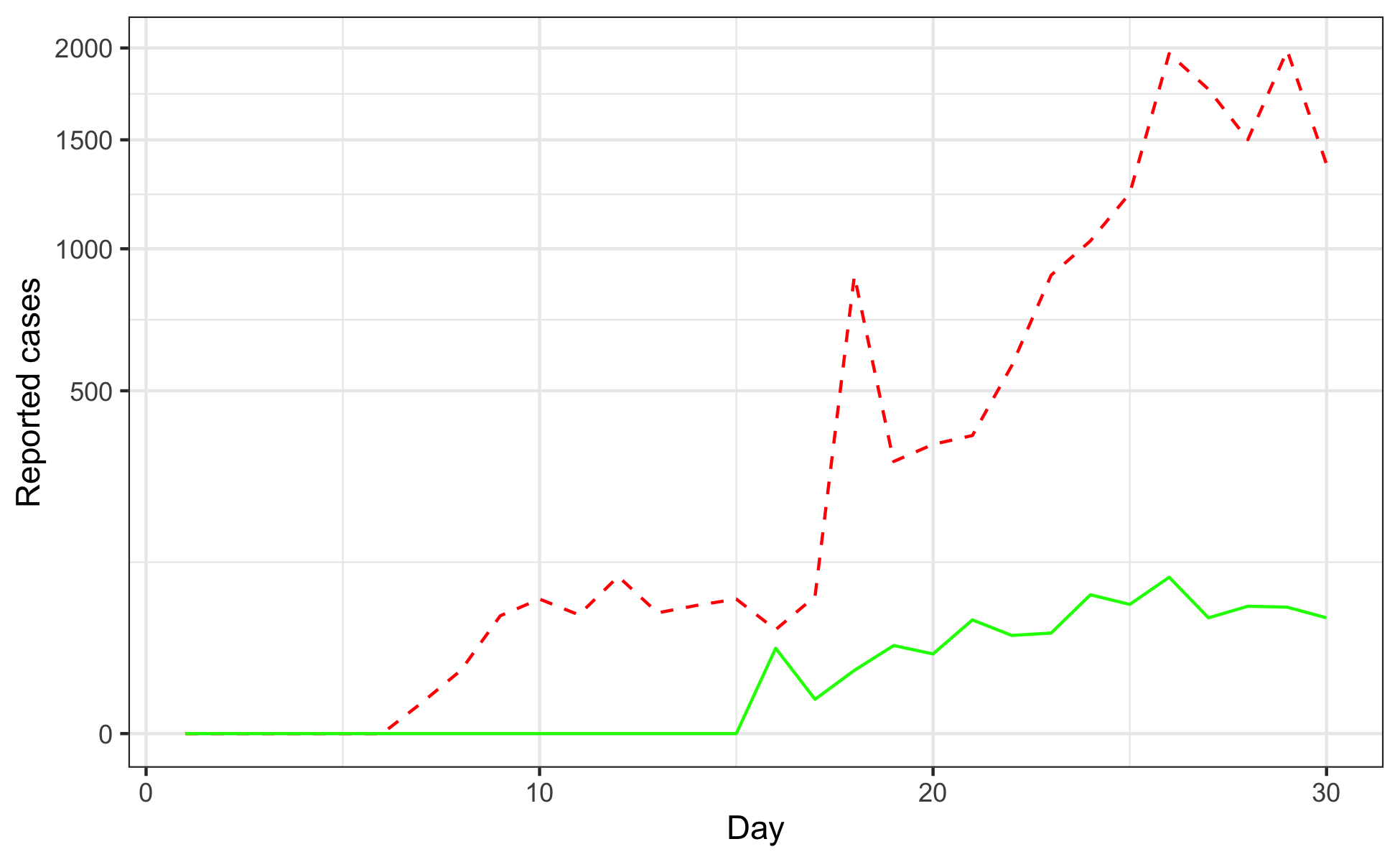} 

}

\caption[Time plot of reported cases in Wuhan (red,dashed) and Huangshi (green,solid)]{Time plot of reported cases in Wuhan (red,dashed) and Huangshi (green,solid).}\label{fig:anomalous-cities-plot}
\end{figure}

\end{knitrout}

Here, we do not show investigations of anomalies that were carried out while developing our model and correcting errors in the data.
The need to correct certain population values described in Section~\ref{sec:li20} was identified by looking to explain why some cities had large anomalies.
Discrepancies between the model and data may be (i) a problem with the model; (ii) an error in the data; (iii) an unavoidable consequence of limitations of the model or data, without being a major flaw in either.
The first task of data analysis is to identify such discrepancies, since that is prerequisite for evaluating what should be learned from them.

%\clearpage

\section{\secSep The metapoppkg R package}
\label{sec:metapoppkg}

Source code reproducing the numerical results in the article and this supplement is available at \url{https://github.com/jifanli/metapop_article}.
The code builds on a software package, \code{metapoppkg}, available at \url{https://github.com/jifanli/metapoppkg}.
This package provides the dataset and models under consideration, as well as some useful data analysis operations.
The documentation and unit tests for \code{metapoppkg} help to make the data analysis extendable: they reduce the overhead for subsequent investigators to adapt the analysis we present with variations to the models, data or statistical methods. Extendable data analysis is discussed by \citet{wheeler23}.

The central component of \code{metapoppkg} is the function \code{li23} builds the model described in Section~\ref{sec:li23}.
The arguments permit specification of the number of spatial units and number of observation times. 
\code{li20} is similar to \code{li23} but aims to replicate the model form of \citet{li20}, as described in \ref{sec:li20}.
\code{R0} evaluates the value of $\Rzero$ for a given set of parameter values.
\code{incidence}, \code{mobility} and \code{population} import the corresponding epidemiological datasets uses for the data analysis.
The \code{metapoppkg} package builds heavily on the \code{spatPomp} package \citep{asfaw23arxiv}, which in turn builds on the \code{pomp} package \citep{king16}.
Generic plotting methods for data, simulations, and diagnostic plots are provided by these packages.
We also use plotting functions designed specifically for the analysis presented here, and made available as the \code{plot\_dist} and \code{plot\_li} functions in \code{metapoppkg}.

\bibliographystyle{apalike}
%\bibliography{../bib-metapop}